\documentclass[sigconf]{acmart}
\renewcommand\footnotetextcopyrightpermission[1]{} % removes footnote with conference information in first column
\AtBeginDocument{%
  }

\usepackage{amsmath,amsfonts}
\usepackage{algorithmic}
\usepackage{graphicx}
\usepackage{textcomp}
\usepackage{xcolor}

\usepackage{xurl}

\usepackage[ruled,vlined]{algorithm2e}
\usepackage{subcaption}
\usepackage{mwe}
\usepackage{hhline}
\usepackage{multirow}
\usepackage{array}
\newcolumntype{?}{!{\vrule width 2pt}}
\usepackage{epigraph}
\usepackage{enumitem}
\usepackage{listings}

\newcommand{\ignore}[1]{}
\newcommand{\profuzz}[1]{ProcessorFuzz}
\newcommand{\difuz}[1]{DIFUZZRTL}
\newcolumntype{?}{!{\vrule width 2pt}}
\newcolumntype{P}[1]{>{\centering\arraybackslash}m{#1}}

 \setcopyright{none}
 
\begin{document}

\title{\profuzz{}: Guiding Processor Fuzzing using\\ Control and Status Registers}

 \author{Sadullah Canakci}
 \email{scanakci@bu.edu}
 \affiliation{%
   \institution{Boston University}
   \state{Boston}
   \country{USA}
 }

 \author{Chathura Rajapaksha}
 \email{chath@bu.edu}
 \affiliation{%
   \institution{Boston University}
   \state{Boston}
   \country{USA}
 }

 \author{Anoop Mysore Nataraja}
 \email{mysanoop@uw.edu}
 \affiliation{%
   \institution{University of Washington}
   \state{Seattle}
   \country{USA}
 }

 \author{Leila Delshadtehrani}
 \email{delshad@bu.edu}
 \affiliation{%
   \institution{Boston University}
   \state{Boston}
   \country{USA}
 }

 \author{Michael Taylor}
 \email{Prof.taylor@gmail.com}
 \affiliation{%
   \institution{University of Washington}
   \state{Seattle}
   \country{USA}
 }
 
 \author{Manuel Egele}
 \email{megele@bu.edu}
 \affiliation{%
   \institution{Boston University}
   \state{Boston}
   \country{USA}
 }

 \author{Ajay Joshi }
 \email{joshi@bu.edu}
 \affiliation{%
   \institution{Boston University}
   \state{Boston}
   \country{USA}
 }


\begin{abstract}

As the complexity of modern processors has increased over the years, developing
effective verification strategies to identify bugs prior to manufacturing has
become critical.
Undiscovered micro-architectural bugs in processors can manifest as severe
security vulnerabilities in the form of side channels, functional bugs, etc.
Inspired by software fuzzing, a technique commonly used for software testing,
multiple recent works use hardware fuzzing for the verification of
Register-Transfer Level (RTL) designs.
However, these works suffer from several limitations such as lack of support for widely-used Hardware Description Languages (HDLs) and misleading
coverage-signals that misidentify ``interesting'' inputs.

Towards overcoming these shortcomings, we present \profuzz{}, a processor fuzzer
that guides the fuzzer with a novel {\em CSR-transition coverage} metric.
\profuzz{} monitors the transitions in Control and Status Registers (CSRs) as
CSRs are in charge of controlling and holding the state of the processor.
Therefore, transitions in CSRs indicate a new processor state, and guiding the
fuzzer based on this feedback enables \profuzz{} to explore new processor states.
\profuzz{} is agnostic to the HDL and does not require any instrumentation in
the processor design.
Thus, it supports a wide range of RTL designs written in different hardware languages.

We evaluated \profuzz{} with three real-world open-source processors -- Rocket,
BOOM, and BlackParrot.
\profuzz{} triggered a set of ground-truth bugs 1.23$\times$ faster (on average)
than DIFUZZRTL.
Moreover, our experiments exposed 8 new bugs across the three RISC-V cores and
one new bug in a reference model.
All nine bugs were confirmed by the developers of the corresponding projects.

\end{abstract}

\maketitle
\pagestyle{plain}
\section{Introduction}\label{sec:Introduction}

As the complexity of processor designs has continuously grown over the years, verification has become one of the most challenging tasks in processor manufacturing.
The state-space of a complex processor is extremely large, while the processor vendors have limited time and resources for verification.
An exhaustive verification (i.e., testing each and every scenario) is an unrealistic goal to achieve, and therefore, a high-quality verification methodology is essential to discover bugs before fabrication.
A timely, pre-silicon bug discovery can circumvent potentially millions-of-dollars of losses~\cite{fdivLosses}.
Otherwise, undiscovered bugs can manifest as severe security vulnerabilities in both proprietary and open-source processors 
such as transient execution vulnerabilities (e.g., Spectre \cite{spectre}, Foreshadow~\cite{foreshadow}), x86’s guest privilege escalation bug~\cite{privEscalateBug}, Intel’s TSX bug \cite{tsx} that breaks KASLR, Intel’s machine check vulnerability~\cite{machineCheck} that enables denial-of-service attacks, and Pentium's FOOF~\cite{foof} and FDIV bugs~\cite{fdiv}.

Broadly, the verification techniques can be divided into two categories - static and dynamic.
Static verification techniques~\cite{chen2011property,mukherjee2015hardware,JasperGold} aim to prove that the implementation is accurate with respect to a specification. 
Due to the well-known state explosion problem~\cite{dessouky2019hardfails} of these techniques, dynamic verification techniques~\cite{fine2003coverage, wagner2005stresstest,tasiran2001functional,nativ2001cost,gal2021automatic,bose2001genetic} are commonly used as part of the processor verification process.  
Dynamic verification involves simulating a Design Under Test (DUT) with a test input and analyzing the behavior of the DUT during or after simulation to identify bugs.
Recent works~\cite{hur2021difuzzrtl, laeufer2018rfuzz, trippel2021fuzzing} demonstrate that Coverage-based Greybox Fuzzing (CGF), a widely-used software testing technique, can be adapted as a dynamic verification technique to identify bugs in a processor design if certain differences between hardware and software are addressed.

Prior works on processor fuzzing mainly focus on addressing \textbf{two major challenges.}
First, code coverage metrics used for \linebreak fuzzing software programs (basic block, branch coverage, etc.) are not well-suited for fuzzing hardware~\cite{tasiran2001functional, hur2021difuzzrtl}.
Second, a bug in a processor design does not result in an observable anomaly (i.e., crash) during testing as opposed to many software programs which can indicate the presence of bugs by throwing memory violation errors or raising exceptions.

\textbf{To address the first challenge}, researchers have introduced a variety of coverage metrics such as multiplexer toggle coverage, register coverage, etc~\cite{laeufer2018rfuzz,hur2021difuzzrtl,muduli2020hyperfuzzing,li2021symbolic} that are tailored for hardware. 
In the context of a processor, the processor is effectively a complex Finite State Machine (FSM) that consists of a large number of states. 
Exploring different states in `processor FSM' is the key to identifying bugs in the processor.
Therefore, hardware-specific coverage metrics mainly aim to guide the fuzzer towards different uncovered `processor FSM' states. 
These metrics take the hardware intrinsic (e.g., wire connections) into account rather than merely the code structure of the hardware.
For instance, DIFUZZRTL~\cite{hur2021difuzzrtl}, a state-of-the-art processor fuzzer, introduces {\em register coverage} metric where the goal is to monitor value changes in registers that control multiplexer selection signals.
The intuition is that a particular value in these registers represents a unique state in the `processor FSM' and guiding the fuzzer based on this feedback explores additional FSM states. 

DIFUZZRTL’s register coverage metric improves on prior works \linebreak ~\cite{laeufer2018rfuzz,moundanos1998abstraction,acharya2015branch} in terms of scalability, efficiency, and precision.
However, we make a key observation that the register coverage can be a highly misleading metric for a processor fuzzer.
Specifically, we find that DIFUZZRTL monitors  many datapath registers which have minimal control over the current FSM state of the processor.
The coverage increase resulting from the datapath registers does not provide meaningful information related to the current FSM state of the processor. 
This results in a scenario where inputs that affect datapath register coverage are incorrectly being classified as `interesting' inputs, which in turn leads to wasted fuzzing time.

\textbf{To address the second challenge}, existing processor fuzzers \linebreak ~\cite{hur2021difuzzrtl, kabylkas2021effective, riscv-torture, riscvdv} adapt differential testing from the software domain to the hardware domain. 
Differential testing in software compares outputs of multiple programs that have the same functional behavior and checks for inconsistencies.
In the hardware domain, the results of an Register Transfer Level (RTL) simulator are compared with those of an Instruction Set Architecture (ISA) simulator.
An RTL simulator is used to simulate the execution of an instruction stream on the detailed microarchitecture implementation of the processor.
The ISA simulator is used to simulate the functional behavior of the processor design and used as a reference model.
A difference in the execution output of RTL simulation and ISA simulation indicates a potential bug in the processor. 

In this work, we present \profuzz{}, a processor fuzzer that implements two novel features.
First, \profuzz{} uses a new coverage metric called {\em CSR-transition coverage} to effectively guide processor fuzzing towards exploring unique processor states.
Specifically, it monitors transitions in Control and Status Registers (CSRs) that form the core of the architecture specifications.
Our intuition is that certain CSRs dictated by ISA readily expose the current `processor FSM' state (e.g., current privilege mode, the event that caused floating mode exception), and thus the transitions in these CSRs signify a new `processor FSM' state.

\profuzz{}'s second feature is that it uses ISA simulation to rapidly determine if a test input is interesting.
Prior works rely on RTL simulation for the same goal, which is time-consuming. 
In fact, this problem gets compounded if the coverage guidance is misleading and results in the execution of repetitive test inputs.
ISA simulation is significantly faster than RTL simulation\footnote{As a reference point, ISA simulation is 79$\times{}$ faster than RTL simulation for an open-source RISC-V based processor (i.e., BOOM~\cite{boom}).}.
Hence, \profuzz{} can efficiently eliminate repetitive test inputs and focus on as many qualitatively distinct test input patterns as possible to expose bugs faster.
Another benefit of this design feature is that \profuzz{} is agnostic to the hardware description language (HDL) used for designing the processor. 
Unlike prior works~\cite{laeufer2018rfuzz,hur2021difuzzrtl}, \profuzz{} does not require any HDL-specific hardware instrumentation because it identifies interesting inputs using ISA simulation.
Hence, processors expressed in different HDLs (VHDL, System Verilog, etc.) can easily utilize \profuzz{} as a verification tool without having to worry about integration issues. 

We evaluate \profuzz{} using a variety of widely-used open-source RISC-V based processors Rocket Core~\cite{rocket}, BOOM~\cite{boom}, and BlackParrot~\cite{blackparrot}.
Here Rocket Core~\cite{rocket} and BOOM~\cite{boom} have been designed using Chisel HDL, while BlackParrot~\cite{blackparrot} has been designed using SystemVerilog. 
In addition, these processors vary in microarchitectural implementations such as their pipeline depths, execution type (i.e., in-order and out-of-order execution), etc.
We compare the bug-finding effectiveness of \profuzz{} against the state-of-the-art register coverage guided DIFUZZRTL. 
On average, for the bugs found by DIFUZZRTL, \profuzz{} triggers bugs 1.23$\times$ faster than DIFUZZRTL.
In addition, \profuzz{} revealed 8 new bugs in widely-used open-source processors and one new bug in a reference model.

In summary, we make the following contributions:

\begin{itemize}
    \item We propose \profuzz{}, a new processor fuzzing mechanism. \profuzz{} uses a novel {\em CSR-transition coverage (CTC)} metric, to effectively guide processor fuzzing towards interesting processor states. 
    \item We propose to use the ISA simulator as part of a coverage feedback mechanism to rapidly identify interesting test inputs, thereby accelerating the bug-finding process.
    \item We demonstrate the practicality of \profuzz{} using 3 different open-sourced RISC-V processors and present eight new bugs identified in those three different processor designs and one new bug in a reference model.
    \item In the spirit of open science and to facilitate reproducibility of our experiments, we will make our source code of \profuzz{} publicly available.
\end{itemize}

\section{Background and Motivation}
\label{sec:background}
\graphicspath{ {./figures/} }

In this section, we first briefly explain coverage-based greybox fuzzing (CGF) for software.
Next, we provide a brief background of how CGF is adapted as a hardware fuzzing method (specifically for processor fuzzing).

\subsection{Coverage-based Greybox Fuzzing}\label{subsec:cgf}
Fuzzing has gained broad adoption in the software community due to its effectiveness in bug discovery, scalability, and practicality~\cite{oss-fuzz,afl,onefuzz}.
Fuzzing is the process of repeatedly running a Program Under Test (PUT) with a large number of random inputs to discover bugs in software.
One of the widely-used fuzzing variants is CGF which utilizes the coverage feedback collected from the PUT at runtime.
In each run of the PUT, CGF records coverage (e.g., basic block coverage, edge coverage, etc.) to determine if the input is `interesting', i.e., whether it leads to increased coverage.
If so, CGF applies a set of mutations to the `interesting' input to generate new inputs which are then fed to the PUT in the next fuzzing rounds.
Here, the intuition is that generating new inputs from coverage increasing ones would cover even more unexplored code. 
CGF instruments the code of the program (either statically or dynamically) with the necessary book-keeping logic to record coverage during the program execution.

\subsection{Adapting CGF for Processor Fuzzing}
Recent works \cite{hur2021difuzzrtl, laeufer2018rfuzz, trippel2021fuzzing} show that CGF can be adapted as a dynamic verification method for hardware including processors.
In this section, we briefly explain two important aspects when adapting CGF to processor fuzzing.

\noindent\textbf{Hardware Execution.} 

In the case of CGF for software, the fuzzing target is a software program that can be directly executed on a host machine with a test input after compilation.
However, hardware (e.g., a processor) is not directly executable on the host machine.
A hardware design is implemented with an RTL abstraction and must be simulated with an RTL simulator to evaluate a test input.
RTL describes the hardware design in terms of data transfer between registers and the logical operations between the registers.
The RTL design is usually expressed with an HDL (e.g., Verilog, VHDL).
The RTL simulator can provide cycle-accurate information regarding the real-time behavior of the RTL design.

\noindent\textbf{Bug Detection.} Most software fuzzers focus on bugs that manifest as memory safety violations such as segmentation faults. 
These types of bugs are relatively easy to detect because they cause an observable anomaly (i.e., crash) in program behavior.  
However, fuzzing to find semantic bugs (e.g., logic errors) is harder than discovering memory violations because defining semantic violations is a highly domain-specific task.
For these types of bugs, researchers proposed differential testing~\cite{sahin2018proteus,martignoni2009testing,brubaker2014using,min2015cross}.
Differential testing compares the output of multiple programs that have the same functionality and checks for inconsistent behaviors.
This approach is used by processor fuzzers~\cite{riscv-torture, riscvdv,hur2021difuzzrtl} as well.  
In particular, the processor fuzzer provides the same input to both the RTL simulator and the reference model.
Here, the reference model is an ISA simulator that mimics the behavior of all the ISA-level operations.
The ISA simulator is a software model of the hardware and does not require any low-level microarchitectural details (e.g., the pipeline depth, buffer size). 
For a given program, it computes the values of architectural registers and memory state after the execution of each instruction.
In contrast, RTL simulator is cycle-accurate and realizes the effect of executed instructions in the microarchitectural level such as the available packets in a buffer or branch prediction result of a conditional branch.   
The hardware fuzzer extracts an execution trace log from both the ISA simulator and the RTL simulator for the same input and cross-checks the traces.
Here, the execution trace contains the final memory states and architectural registers.
Any mismatch in the traces is considered a potential bug in the processor and is marked for further investigation by the verification engineer.

\subsection{\difuz{}'s Register Coverage}\label{difuz-motiv}
\begin{figure*}[]
    \centering
    \includegraphics[width=1\textwidth]{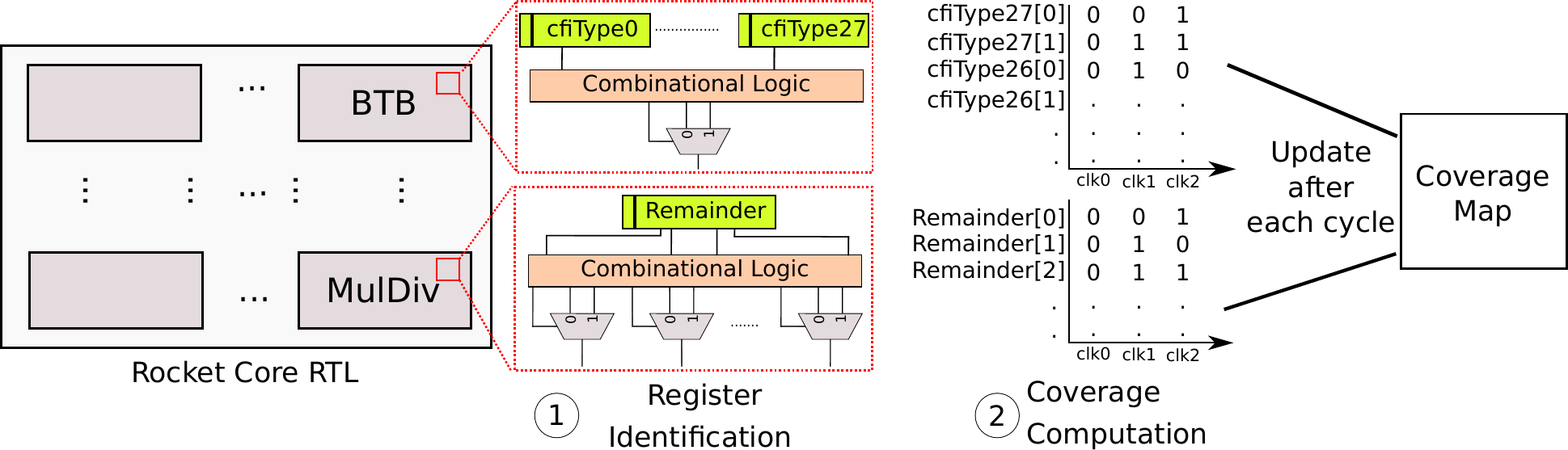}
    \caption{Overview of \difuz{}'s coverage feedback strategy. \difuz{} detect multiplexer select signals in the RTL design and trace them back to find the registers that affect these select signals (\textcircled{1}). These registers are then hashed together to represent the FSM state of each module(\textcircled{2}). \difuz{} detects data-path registers such as \texttt{remainder} from Rocket core MulDiv module which increase the state space of the module significantly.}
    \label{fig:motivation}
\end{figure*}

\difuz{}~\cite{hur2021difuzzrtl}, the current state-of-the-art hardware fuzzer, uses this CGF approach to identify bugs in processors.
\difuz{} proposes a feedback strategy that aims to capture FSM state transitions during RTL simulation.
The strategy follows a two-stage approach as depicted in Figure~\ref{fig:motivation}.
In stage \textcircled{1}, it performs static analysis to identify a small set of registers in each RTL module and instruments the RTL with necessary hardware logic to record register coverage at simulation time. 
At a high level, \difuz{} monitors a register if its value is directly or indirectly used to control a multiplexer selection signal.
\difuz{} creates a circuit graph of the RTL design where nodes and edges of this graph represent circuit elements (e.g., multiplexers, wires, ports, registers) and connections, respectively.
Then, it recursively performs a backward data-flow analysis for each multiplexer's selection signal and identifies any register in the traversed path.
In stage \textcircled{2}, \difuz{} monitors value changes in the identified registers during the RTL simulation. 
For each clock cycle, \difuz{} hashes all the values in the identified registers into a coverage map to represent the current FSM state. 
If a new hash value is observed, \difuz{} increases register coverage to signify that the current test is interesting for further mutations.

\difuz{}'s register coverage improves prior work~\cite{laeufer2018rfuzz, li2021symbolic} in terms of scalability, efficiency, and precision. 
However, using register coverage metric for hardware fuzzing can be highly misleading.
At a high level, we observe that a subset of registers leads to misleading coverage increase, and therefore, misguides the hardware fuzzer.
We provide more details using an example (illustrated in Figure~\ref{fig:motivation}) from the open-source RISC-V-based Rocket Core~\cite{rocket}.

\begin{figure}[]
        \centering
        \begin{subfigure}[b]{0.49\columnwidth}
            \centering
            \includegraphics[width=\columnwidth]{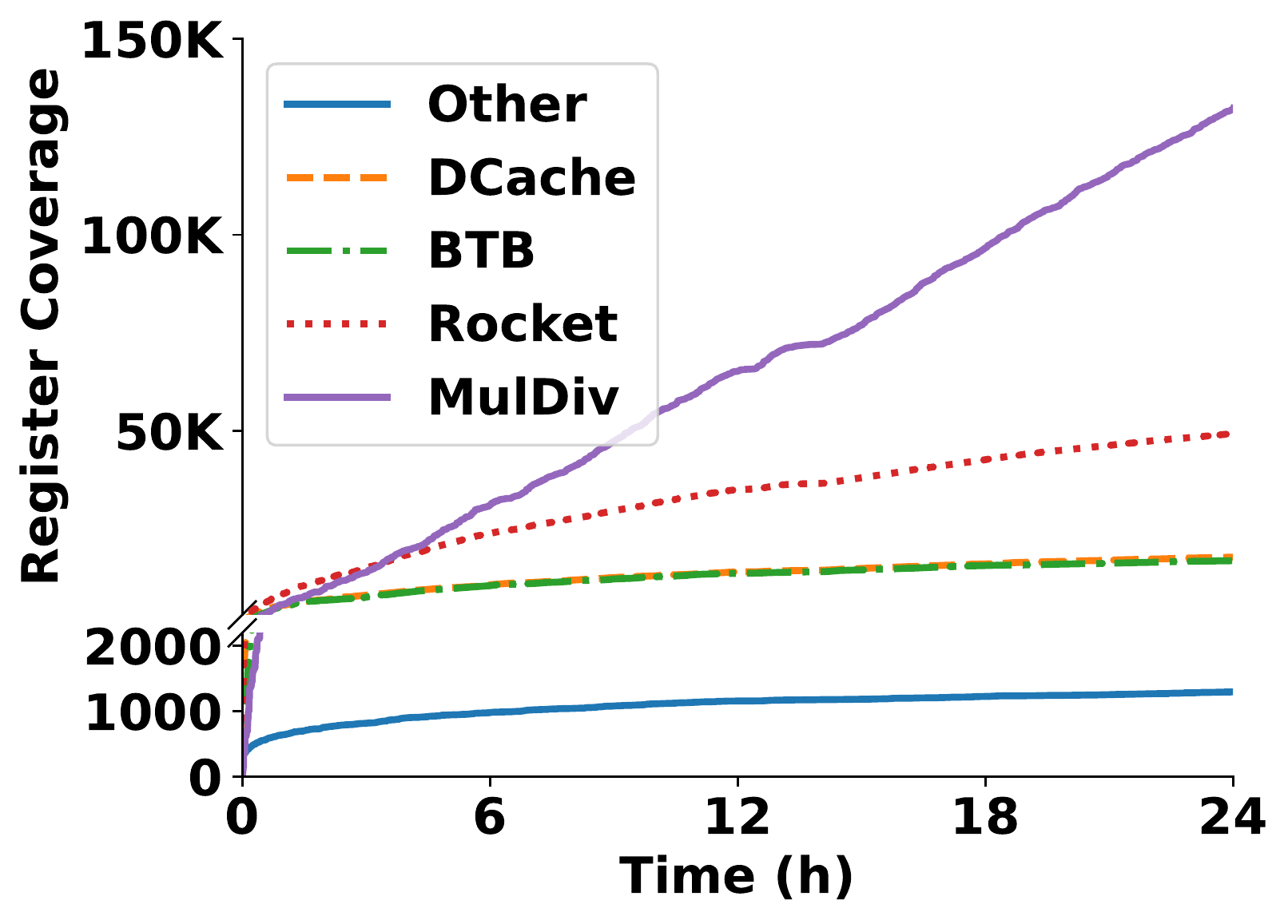}
            \caption[Network2]%
            {{\small Rocket core}}    
            \label{fig:rocket-break}
        \end{subfigure}
        %\vspace{-0.5cm}
        \begin{subfigure}[b]{0.49\columnwidth}
            \centering
            \includegraphics[width=\columnwidth]{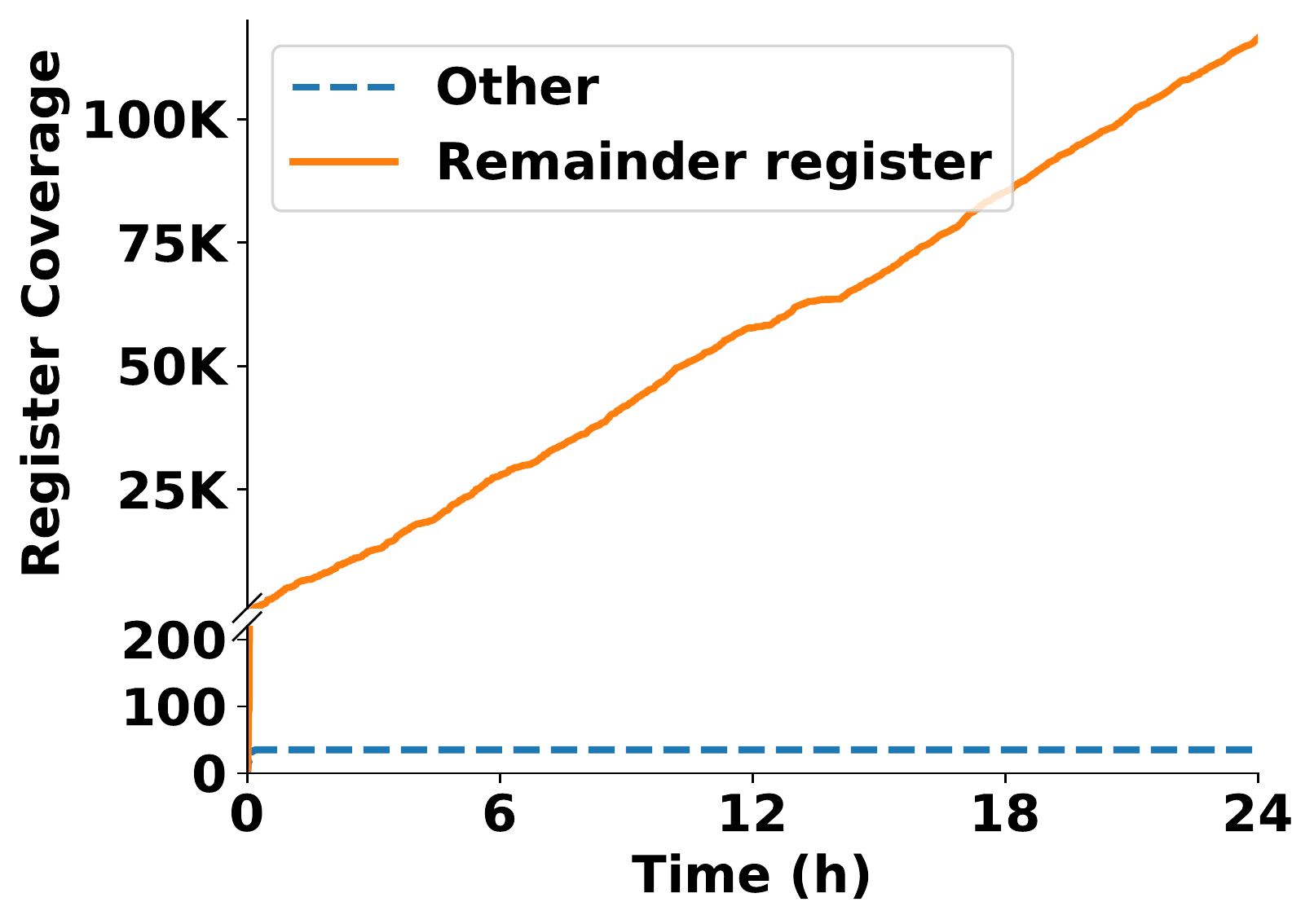}
            \caption[Network2]%
            {{\small MulDiv module}}    
            \label{fig:muldiv-break}
        \end{subfigure}
        %\vspace{-0.5cm}
        \caption{\difuz{}'s register coverage breakup for Rocket core over time.}
        \label{fig:regcov_break}
    \end{figure}

In particular, in the multiplication unit of Rocket Core, there is a 130-bit \texttt{remainder} register in the MulDiv module that indirectly controls 98 mux selection signals.
Therefore, \difuz{} identifies this register to monitor during fuzzing.~\footnote{\difuz{} applies some optimizations to reduce search space. As one of their optimizations, it is able to track only a subset of bits of a register and therefore; ultimately tracks 98-bit in \texttt{remainder} register.} 
The change in the value of \texttt{remainder} results in an increase in coverage.
In Figure \ref{fig:regcov_break}, we demonstrate the coverage increase resulted from the \texttt{remainder} register during a 24-hour fuzzing session.
First, in Figure \ref{fig:rocket-break}, we depict the coverage progress of different modules in the Rocket core.
Clearly, the MulDiv module (multiplication unit of Rocket core) dominates the module-wise register coverage. 
62\% of overall register coverage results from the MulDiv module at the end of 24-hours.
Figure \ref{fig:muldiv-break} further  shows the contribution of the \texttt{remainder} register to the coverage increase in the MulDiv module.
Compared to all other registers in the MulDiv module, \texttt{remainder} register is clearly major factor that causes increase in register coverage.
Indeed, our further analysis showed that the multiplication of two different numbers (see code snippet in List~\ref{mylist0}) increases the register coverage (i.e., explores a new state) even after 2M iterations. 

\definecolor{mGreen}{rgb}{0,0.6,0}
\definecolor{mGray}{rgb}{0.5,0.5,0.5}
\definecolor{mPurple}{rgb}{0.58,0,0.82}
\definecolor{backgroundColour}{rgb}{0.95,0.95,0.92}

\lstset{
%backgroundcolor=\color{backgroundColour},
   commentstyle=\color{mGreen},
   %identifierstyle=\color{blue},
   %keywordstyle=\color{red},
   %numberstyle=\tiny\color{mGray},
   %stringstyle=\color{mPurple},
   basicstyle=\footnotesize{} , %
   breakatwhitespace=false,
   breaklines=true,
   captionpos=b,
   keepspaces=true,
  % numbers=left,
   numbersep=6pt,
   showspaces=false,
   showstringspaces=false,
   showtabs=false,
   tabsize=3,
   caption={Code snippet for testing multiplication unit.},
   label=mylist0,
   language=C,
   numbers=left,xleftmargin=2em,frame=single,framexleftmargin=1.5em
}

\begin{lstlisting}
void main() {
  unsigned int num1, num2, res;
  for (int i = 0; i < 2000000; i++){
    num1 = i; num2 = i+1;
    res = num1 * num2; 
  }
}
\end{lstlisting}

Broadly, as pointed out with the above  example, \difuz{} monitors and uses coverage information from registers even if they are mostly involved in datapath-related operations and have minimal control over the current FSM state of the hardware.
Unfortunately, data-path registers (e.g., \texttt{remainder}) increase search space significantly, yet the coverage increase resulting from data-path registers indeed does not provide  meaningful information to the fuzzer related to the current hardware state.
Therefore, it is not interesting to keep an input for further mutations if it increases coverage based on data-path registers. 
In our work, we present a new coverage metric that aims to tackle this problem.
\section{\profuzz{}}
\label{sec:methodology}

In this section, we present the design of \profuzz{}, a fuzzing mechanism tailored for processors.
We first provide a high-level overview of the different stages of \profuzz{}. 
Then,  we outline our reasoning for using ISA simulation instead of RTL simulation to evaluate coverage.
Finally, we explain the details of our CSR transition coverage metric and how \profuzz{} uses this metric to guide the fuzzing procedure.

\subsection{Design Overview}
\begin{figure*}[]
    \centering
    \includegraphics[width=1\textwidth]{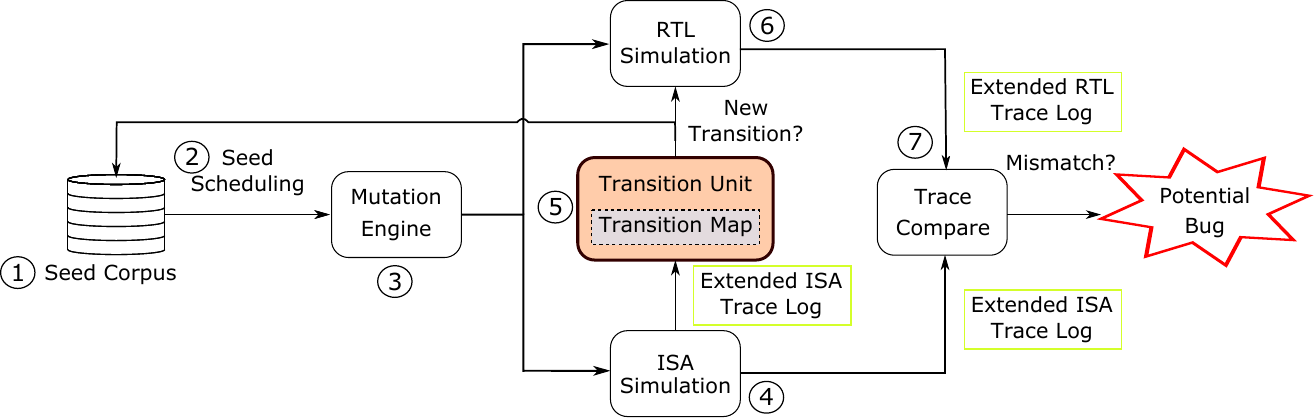}
    \caption{\profuzz{} Design: \profuzz{} runs the ISA simulator with an input generated by the mutation engine and outputs an extended ISA trace log that contains CSR values. The transition unit extracts CSR transitions, determines if a transition is new by checking the transition map, and stores new ones in the transition map. \profuzz{} runs the RTL simulation only with interesting inputs and creates an RTL trace log to be compared with the ISA log for bug detection.}
    \label{fig:isa-sim-fuzz}
    %, and determines interesting tests by checking the transition map.
\end{figure*}

We illustrate the design overview of \profuzz{} in Figure \ref{fig:isa-sim-fuzz}.
In stage \textbf{(1)}, \profuzz{} is provided with an empty seed corpus. 
It populates the seed corpus by generating a set of random test inputs in the form of assembly programs that conforms to the target ISA.
Next, \profuzz{} chooses a test input from the seed corpus in stage \textbf{(2)} and subsequently applies a set of mutations (such as removing instructions, appending instructions, or replacing instructions) on the chosen input in stage \textbf{(3)}.
For these three stages, \profuzz{} uses the same methods applied by a prior work~\cite{ hur2021difuzzrtl}.
In stage \textbf{(4)}, \profuzz{} runs an ISA simulator with one of the mutated inputs and generates an extended ISA trace log.
A typical trace log generated by the ISA simulator contains (for each executed instruction) a program counter, the disassembled instruction, current privilege mode, and a write-back value as detailed in Section~\ref{sec:background}.
The extended ISA trace log additionally includes the value of CSRs for each executed instruction.
The Transition Unit (TU) receives the ISA trace log in stage \textbf{(5)}.
The TU extracts the transitions that occur in the CSRs. 
Each observed transition is cross-checked against the Transition Map (TM). 
The TM is initially empty and populated with unique CSR transitions during the fuzzing session.
If the observed transition is not present in the TM, it is classified as a unique transition and added to the TM.
In case the current test input triggers at least one new transition, the input is deemed interesting and added to the seed corpus for further mutations. 
If, however, there are no new transitions triggered, the input is discarded. 
In stage \textbf{(6)}, \profuzz{} runs the RTL simulation of the target processor with the mutated input only if the input is determined as interesting.
The RTL simulation also generates an extended RTL trace log similar to the extended ISA trace log.
The extended RTL trace log contains the same information as the extended trace log.
The ISA trace log and the RTL trace log are compared in stage \textbf{(7)}.
Any mismatch between the logs signifies a potential bug that needs to be confirmed by a verification engineer usually by manual inspection. 

\begin{figure}[]
    \centering
    \includegraphics{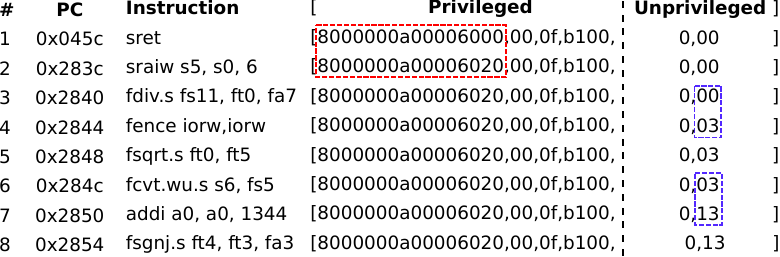}
    \caption{Extended trace log generated by the ISA simulator. The values (in hexadecimal) of a subset of CSRs in Table \ref{tab:csrlist} in Appendix are included within the square brackets in the given order; \texttt{mstatus}, \texttt{mcause}, \texttt{scause}, \texttt{medeleg},
    \texttt{frm}, and \texttt{fflags}. Transitions are color coded; red and blue for \texttt{mstatus} and \texttt{fflags} CSR transitions, respectively.}
    \label{fig:tracelog}
\end{figure}

\subsection{Feedback from the ISA Simulation} 
\label{isa-sim-fed}
One design feature of \profuzz{} is that it relies on the ISA simulation to determine if a test input is interesting as opposed to prior works that rely on the RTL simulation.
Specifically, \profuzz{} runs the ISA simulator with each input obtained from the mutation engine and collects necessary feedback (i.e., CSR transitions which we detail in the following subsections) from the simulator. 
\profuzz{} later processes the collected feedback to determine if the input should be ignored or used by the RTL simulator.

We use the ISA simulator to capture the CSR transitions for two main reasons.
First, ISA simulators are generally much faster in executing a given program in comparison to executing that program on a processor using the RTL simulation. 
For instance, we observed that the RISC-V Spike ISA simulator is, on average 79$\times$ faster than the RTL simulation of the RISC-V BOOM processor.
This speedup provides a considerable advantage as \profuzz{} can then quickly identify if a test input is interesting without performing the slow RTL simulation.
Eliminating inputs with similar characteristics help \profuzz{} to achieve faster bug discovery times as shown in Section~\ref{sec:evaluation}.
Indeed, \profuzz{} discovered all the bugs found by the existing processor fuzzer (i.e., \difuz{}). 

Second is the reduced effort needed to instrument the simulator. 
A simulator needs to be instrumented to generate an extended trace log with the selected CSRs. 
An ISA simulator can be easily instrumented by extending the already available trace logic with the selected CSRs.
The same instrumented ISA simulator can be used to fuzz any processor design as long as it has been designed for the same ISA target. 
In contrast, instrumenting RTL designs for tracking the coverage metrics requires extensive effort.
Moreover, instrumentation in one HDL does not readily translate to other HDLs.
Additionally, as shown in Section~\ref{sec:evaluation}, \profuzz{} incurs limited instrumentation overhead during fuzzing (only <1\% in ISA simulator) as opposed to prior works~\cite{tyagi2022thehuzz} that instrument processor RTL and result in higher runtime overheads (e.g., ~71\% by TheHuzz~\cite{tyagi2022thehuzz} and \%97 by RFUZZ~\cite{laeufer2018rfuzz}).

\subsection{CSR-transition Coverage} \label{subsec:csrcov}

\subsubsection{Description of the Metric}
As described in Section~\ref{difuz-motiv}, DIFUZZRTL's register coverage technique monitors many datapath registers (e.g., \texttt{remainder} register) to determine the current FSM state, which leads to large state space.
Hence, guiding the fuzzing procedure with DIFUZZRTL's register coverage metric can be highly misleading when fuzzing processors.
To test the processor with as many qualitatively distinct input patterns as possible, we propose a novel CSR transition-based coverage metric.

CSRs are system registers in an ISA specification.
These registers are used to control (e.g., delegated exceptions) or hold information (e.g., state of the floating-point unit) about the current architectural state of the processor.
Our intuition for using CSRs is as follows.
A processor is a complex FSM where CSRs have direct control over the current processor state.  
Architectural state of the processor (held in the register file and status registers) represents the state of a program running in the processor.
A value change in a CSR often signifies an architectural state change such as a value change in a CSR that stores exception code or privilege level.
Therefore, \profuzz{} aims to realize the current state of the processor by monitoring transitions in CSRs to guide the fuzzer towards interesting processor states.

\begin{figure}[!b]
    \centering
    \includegraphics[]{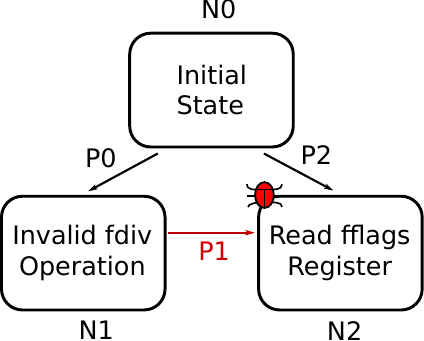}
    \caption{Abstract state diagram for triggering Bug 2 listed in Table \ref{tab:bugDescription}}
    \label{fig:bug_state}
\end{figure}

CSRs are part of both an ISA simulator and the RTL design of a processor.
Hence, CSR transitions can be extracted either from the ISA simulation or the RTL simulation.
As detailed before, \profuzz{} uses the ISA simulator to capture CSR transitions. 
Specifically, to extract a CSR transition, \profuzz{} monitors the CSR values resulting from the execution of the previous and current instructions and checks if they differ. 
If so, \profuzz{} uses the transition to determine if the input is interesting as detailed in the following subsections.
Here, we provide a concrete example to illustrate how \profuzz{} identifies a CSR transition in the ISA trace log.
Consider the extended ISA trace log shown in Figure \ref{fig:tracelog}. 
The CSR value changes after execution of the \textit{sret} instruction shown in Line 1, which can be seen by comparing the entries in Line 1 and Line 2 of the `Privileged' column.
Specifically, we observe a CSR-transition in \texttt{mstatus} CSR from $0x8000000a00006000$ to $0x8000000a00006020$ as highlighted in red in Figure~\ref{fig:tracelog}.
Overall, from Figure \ref{fig:tracelog}, we represent the CSR transition caused by \textit{sret} instruction as ($S_0, S_1$) = (8000000a00006000000fb100000, 8000000a00006020000fb100000), where $S_0$ and $S_1$ are defined as the concatenated CSR values before and after the transition, respectively..

\subsubsection{Why Transitions Instead of Values?}

DIFUZZRTL determines the current processor state based on the register coverage as detailed in Section~\ref{difuz-motiv}.
For each newly covered FSM state, DIFUZZRTL's register coverage only stores the current state of the processor and does not consider the previous state.
Unfortunately, this design choice can lead to important test inputs being discarded by the fuzzer and the fuzzer can potentially miss out on the discovery of a bug. 
We illustrate this in detail below. 
Figure \ref{fig:bug_state} represents a subset of the abstract states associated with a real-world bug (Bug 2 in Table \ref{tab:bugDescription}) that we identified in an open-source RISC-V processor. 

In the figure, the processor starts out in the N0 state. 
The bug triggers in the N2 state only if the previous state is N1.
It does not trigger when the previous state is N0.
During a coverage-guided fuzzing session, if both N1 (through P0 transition) and N2 (through P2 transition) are covered individually, there will not be a coverage increase for the denoted P1 state transition. 
And so, the unique P1 transition is not particularly driven towards. 
Thus, the fuzzing session fails to trigger the bug.
Contrarily, by monitoring transitions, we can detect P1 as a new transition even though N1 and N2 states are already covered.  
Overall, we monitor new transitions in CSRs rather than just identifying unique CSR values to improve the sensitivity of the feedback metric. 
Indeed, our rationale is similar to widely-used software fuzzers~\cite{afl,libfuzz}'s rationale that  monitors edges in a program instead of basic blocks.
We provide the details on how \profuzz{} extracts CSR transitions in the next subsection.

\subsubsection{CSR Selection Criteria} \label{csr-selection}
An ISA specification usually specifies a large number of CSRs\footnote{As a reference point, RISC-V ISA defines up to 4096 CSRs}.
Monitoring all available CSRs for transitions can mislead the fuzzer (as we show in Section~\ref{sec:evaluation}) because not all CSRs provide distinctive information regarding the current processor state. 
As an example, consider \texttt{instret} CSR that holds the total number of retired instructions.
Considering this CSR results in a scenario where each committed instruction by the processor results in a CSR transition.
Effectively, \profuzz{} would identify any test input as interesting since the \texttt{instret} CSR causes a transition after each committed instruction.  
However, a test would rarely result in a bug because of a change in committed instruction count.
To aid \profuzz{} in determining qualitatively different inputs,  we introduce the following two criteria when selecting the CSRs that \profuzz{} monitors transitions. 
First, we select CSRs that contain status information about the processor (criteria C1).
These CSRs are important because they directly reveal the current status of the processor.
As an example, we select a CSR that stores the cause for an exception taken by the processor (e.g., \texttt{mstatus}).
If a test case results in an exception, \profuzz{} analyzes the cause and differentiates it from another test case that has a different exception reason (e.g., misaligned load/store attempt or access faults due to unauthorized privilege mode).
Second, we select any CSR that is used to set a certain configuration in the processor (criteria C2).
Here, we aim to realize if the processor behaves as expected under different configurations.
For instance, the value of \texttt{medeleg} can be changed to determine which traps can be delegated to lower privilege levels (e.g., the load access fault handled in supervisor mode instead of machine mode).
This way, \profuzz{} aims to realize if processor designs can perform correctly under different configurations (e.g., different exception delegations) for a particular processor status (e.g., an exception).
In Table~\ref{tab:csrlist} in Appendix, we list all the CSRs in the RISC-V ISA that we used for identifying transitions in the current implementation of \profuzz{} based on the aforementioned two criteria C1 and C2. 
We also provide all the CSRs that we excluded (e.g., \texttt{instret}) along with details why they are not considered as part of \profuzz{}'s current design (Table \ref{tab:csrexcludelist} in Appendix).

Apart from these two criteria, CSR selection can be further limited depending on the features supported by the target processor or the desired scope of verification.
For example, if the target processor does not support interrupts within the testing framework, any CSRs related to the configuration or status of interrupts can be excluded. 
Similarly, if we only want to verify the functionality of the floating-point unit in the processor, only floating-point CSRs can be monitored to identify transitions.
We quantitatively demonstrate this capability of \profuzz{} in Section \ref{groundtruth}.

 \begin{figure*}[t]
    \centering
    \includegraphics[width=0.9\textwidth]{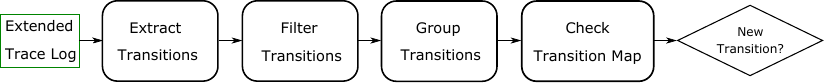}
    \caption{Workflow of the \profuzz{} transition unit.}
    \label{fig:transitionextractor}
\end{figure*}

\subsection{Transition Unit} 
As shown in Figure \ref{fig:isa-sim-fuzz}, the TU takes an extended ISA trace log as input and communicates with the TM to output whether the trace log contains any new transitions.
We describe the complete workflow of the TU in Figure \ref{fig:transitionextractor}.  
As a first step, the TU extracts all CSR transitions in the trace log based on the description in Section~\ref{subsec:csrcov}.
Then, \profuzz{} applies a filter to remove unnecessary transitions. 
Next, the TU groups the transitions to reduce the state space.
We describe how the TU filters  and groups the transitions in the rest of this subsection. 

\noindent \textbf{Filtering Transitions.} 
We note that the number of possible CSR transitions can be large depending on the cumulative width of the selected CSRs.
However, not all CSR transitions represent interesting architectural state changes that are relevant for testing processors. 
For instance, a test program running on the target processor can write to a CSR that contains processor status, e.g. \texttt{mstatus} CSR in RISC-V ISA.
This could get identified as a new CSR transition. 
If the write operation is legal, the processor continues the execution of the program and eventually overwrites the CSR with the updated status.
Overall, the type of transitions that occur from writes to status CSRs do not affect the architectural state of the processor.
Thus, \profuzz{} filters out transitions that occur from explicit writes to status CSRs.

\noindent \textbf{Grouping Transitions.} 
\profuzz{} provides the flexibility to customize the CSR-transition coverage metric to be suitable for verifying  different Architectural Units (AUs) individually. 
Specifically, \profuzz{} allows a designer to group CSR transitions of AUs, thereby considering them as independent events.
Grouping transitions improves the exploration of CSR transitions within each group. 
As a result, the fuzzer is able to generate tests targeted towards individual AUs and verify them thoroughly.  
This is a useful feature for a verification engineer as AUs in a processor can be individually verified as an initial step of verification.
For example, privileged and unprivileged architectures in a RISC-V processor can be verified individually by grouping transitions as shown in Figure \ref{fig:tracelog}. 
Identifying and fixing the bugs in each AU before fuzzing the processor as a whole can reduce the overall verification effort. 

\noindent \textbf{Transition Map} 
\profuzz{} maintains a transition map to store CSR-transitions. 
Each transition is stored in the map as a tuple: $(I_m, S_0, S_1)$ where $I_m$ is the mnemonic of the instruction whose execution resulted in the CSR transition.
$S_0$ and $S_1$ are CSR values before and after the transition as defined in subsection \ref{subsec:csrcov}.
Revisiting the same example given in subsection \ref{subsec:csrcov}, privileged CSR-transition caused by \textit{sret} instruction can be represented as (\textit{sret}, 8000000a00006000000fb1000000, 8000000a00006020000fb1000000).
Similarly, \profuzz{} converts the unprivileged CSR-transition in lines 3 and 4 in Figure \ref{fig:tracelog} to (fdiv.s, 0000, 0003). 

We include instruction mnemonic in the aforementioned tuple because the same transition can be triggered by different instructions. 
For example, both floating-point division and floating-point square-root instructions can trigger the same transition in fflags CSR in RISC-V ISA due to invalid operations. 
Nevertheless, only the invalid operation of floating-point division instruction might contain a bug.
Therefore, we tag each transition with the mnemonic of the instruction that triggered it to uniquely identify transitions triggered by different instructions. 
Only the mnemonic of the instructions is included to ignore repetitive transitions that get triggered by different operands of the same instruction.

Once tuples are created, the map is queried to check whether the detected transition is new or a duplicate. 
Tuples that are identified to contain new transitions are added to the map while marking the current test input as interesting.
The transition map is empty at the beginning of a fuzzing session and maintained throughout the session.

\subsection{RTL Simulation and Trace Comparison}
If the TU determines that the current input results in a unique CSR transition, \profuzz{} launches the RTL simulation and generates the extended RTL trace log.
\profuzz{} then compares the extended RTL trace log with the extended ISA trace log.
Any difference between these logs signifies a potential bug in the processor design and needs to be investigated further by a verification engineer.
In case the input does not result in a unique transition, \profuzz{} discards the input and proceeds to the next fuzzing iteration.
\section{Evaluation}
\label{sec:evaluation}

In this section, we evaluate the effectiveness of \profuzz{} using real-world processor designs.
First, we provide the details of our evaluation setup.
Then, we assess the bug-finding capability of \profuzz{} using ground-truth bugs and compare \profuzz{}'s performance against \difuz{}.
Specifically, we analyze if \profuzz{} can expose the same set of bugs reported by DIFUZZRTL in a more efficient way.
Finally, we describe the list of new real-world bugs that \profuzz{} identified along with the severity of the bugs. 

\subsection{Evaluation Setup}
\subsubsection{Implementation Details} \profuzz{} has two main implementation steps; generation of an extended trace log using the ISA simulator and building the TU (see Figure~\ref{fig:isa-sim-fuzz}). 
For the former, we extended SPIKE~\cite{spike} open-source ISA simulator to store the values of monitored CSRs (see Table~\ref{tab:csrlist} in Appendix) for each executed instruction during the ISA simulation.
The instrumentation overhead of SPIKE is 0.4\% in terms of lines of C++ code, while the runtime overhead is 0.15\%.
The TU is implemented as a Python library.
For the RTL simulation of all processors designs, we used Verilator~\cite{verilator},  an open-source RTL simulator.
We used the same mutation engine (see Figure~\ref{fig:isa-sim-fuzz}) as provided by DIFUZZRTL's open-source repository.
Using the same engine is important since our goal is to compare two coverage feedback mechanisms (i.e., register coverage and CSR-transition coverage) rather than input generation mechanisms.
We separated transitions belonging to \texttt{frm} and \texttt{fflags} to separate floating-point operations from the rest of the CSRs. 

\subsubsection{Processor Designs}
In our evaluation, we use three real-world open-source processors designed using the open-standard RISC-V ISA.

\noindent\textbf{RISC-V Rocket Core.} Rocket core is an open-source, general-purpose, in-order, RISC-V processor core that can be generated using the Rocket Chip SoC Generator framework~\cite{rocket}.
Rocket core is designed in Chisel HDL~\cite{chisel}, and is shown to integrate well with custom hardware accelerators.
Rocket core has been taped out multiple times~\cite{rocket} and is capable of booting Linux. 
Essentially, it is well-tested.
We used Spike~\cite{spike} as a reference model to verify the correctness during fuzzing. 
The commit version of the Rocket core that we used is \texttt{148d5d2}.

\noindent\textbf{RISC-V BOOM Core. }
BOOM~\cite{boom} core can also be generated from the same Rocket Chip SoC Generator framework~\cite{rocket} and is also designed in Chisel HDL.
BOOM is an out-of-order, superscalar RISC-V processor core and capable of booting Linux. BOOM has also been taped out~\cite{boomtapeout}.
We used Spike ISA simulator to verify the correctness during fuzzing. 
The commit version of the BOOM core that we used is \texttt{148d5d2}.

\noindent\textbf{RISC-V BlackParrot Core. }
BlackParrot~\cite{blackparrot} is an open-source 64-bit RISC-V core, designed in the industry-standard SystemVerilog HDL. 
BlackParrot is an ideal candidate for hosting accelerator fabrics and for hardware research owing to its tiny, modular, and friendly design approach.
BlackParrot is silicon-validated and is in active development.
We used Dromajo~\cite{dromajo} as a reference model to expose the bugs in BlackParrot for the \texttt{bc3b48b} commit version. 

\subsubsection{Settings}\label{sec:settings}
We compared \profuzz{} with two different settings of DIFUZZRTL.
The first setting is \texttt{no-cov-difuzzrtl} where DIFUZZRTL fuzzing framework is used without any coverage guidance (i.e., as a blackbox fuzzer).
For all the cores that we evaluated, we successfully used this setting as a comparison point.
The second setting is \texttt{reg-cov-difuzzrtl} where DIFUZZRTL fuzzing framework relies on register coverage as a guidance mechanism.
While this setting is applicable to Rocket and BOOM Cores, it is not the case for BlackParrot Core. 
This is because DIFUZZRTL's register coverage passes do not support SystemVerilog. 
They are tailored for FIRRTL~\cite{firrtl}, an intermediate representation (IR) used by Chisel HDL, which is used to design Rocket and BOOM cores.
We tried to convert SystemVerilog to FIRRTL using an open-source tool (i.e., Yosys~\cite{Yosys}), and apply \difuz{}'s register coverage passes. 
However, we observed several issues during this conversion due to the limited support for SystemVerilog to FIRTTL conversion and thus failed to instrument BlackParrot.  
In our experiments, we used \difuz{} as the sole comparison point since it shows clear benefits over previous processor fuzzing frameworks as well as its open-source nature.
Also, for each setting, we reported Time-to-Exposures (TTE) which is defined as the total elapsed time from the starting of the fuzzing session until the bug is exposed.

\subsubsection{Infrastructure}
All the experiments based on ISA and the RTL simulations were conducted on server nodes with Intel\textsuperscript{\textregistered}Xeon\textsuperscript{\textregistered}E5-2670 CPUs and CentOS Linux 7 as the operating system. 
We fuzzed each processor design 10 times for each setting and allocated 48 hours (2 days) of time limit for each fuzzing instance.
For each fuzzing instance, we dedicated two cores and 8GB of memory.
In total, it took 4320 CPU hours to conduct all the experiments detailed in the following sections.

\subsection{Ground-truth Bugs}
\label{groundtruth}

As discussed by many prior works~\cite{manes2019art, klees2018evaluating}, the bug-finding capability of a fuzzer is the ultimate litmus test for a fuzzer.
While there exist several fuzzing benchmarks for software programs~\cite{li2021unifuzz,hazimeh2020magma}, this is not the case for processors.
Therefore, we relied on a set of bugs (in total six bugs) previously reported by DIFUZZRTL for BOOM processor to evaluate the bug-finding capability of \profuzz{} and perform a head-to-head comparison with DIFUZZRTL.
Overall, our evaluation aims to demonstrate that \profuzz{} can guide the fuzzer efficiently to discover ground-truth bugs thanks to the CSR-transition feedback obtained using the ISA simulation.
In summary, \profuzz{} was able to discover all the ground truth bugs and achieved lower TTE compared to \difuz{}.

In Table~\ref{tab:tab1}, we report the TTE of bugs in seconds for three different settings in 2nd-4th columns; \texttt{no-cov-difuzzrtl}, \texttt{reg-cov\--difuzzrtl}, and \profuzz{} for the BOOM processor core. 
We also provide the achieved speedups by \profuzz{} over \texttt{no-cov\--difuzzrtl}, and \texttt{reg-cov-difuzzrtl}.
For \profuzz{}, we provide results for three different configurations; \texttt{selected}, \texttt{fp-csr}, and \texttt{all-csr}.
These configurations differ in the CSRs that \profuzz{} monitors during fuzzing.
Specifically, \texttt{selected} configuration of \profuzz{ } uses the CSRs in Table \ref{tab:csrlist} in Appendix for transition extraction based on the criteria that we detailed in Section~\ref{csr-selection}.
\texttt{all-csr} configuration monitors all implemented CSRs in  the BOOM core.
Here, by using \texttt{all-csr} configuration, we aim to present that \profuzz{} can be effectively guided towards bugs by eliminating certain CSRs that do not assist fuzzing towards exploring bugs (e.g., \texttt{minstret} that repeatedly changes after an instruction retires). 
Finally, \texttt{fp-csr} configuration uses only the floating-point CSRs (unprivileged CSRs in Table \ref{tab:csrlist} in Appendix).
The aim of this experiment is to show that \profuzz{} can focus on certain parts of processors by selecting a subset of CSRs (e.g., floating point unit). 
Overall, \profuzz{} \texttt{selected} configuration and \difuz{} discovered five out of six bugs reported in the DIFUZZRTL within the fuzzing time limit in our experiments.
Unfortunately, we could not detect \#504 with any of the settings.
In summary, \profuzz{} (\texttt{selected}) achieved, on average, 1.21$\times{}$ (up to 2.1$\times$) and 1.23$\times{}$ (up to 2.32$\times$) speedups over \texttt{no-cov-difuzzrtl} and \texttt{reg-cov-difuzzrtl}, respectively. 
\texttt{no-cov-difuzzrtl} \linebreak performed slightly better than \texttt{reg\-cov-difuzzrtl}.

We included \texttt{fp-csr} configuration to  demonstrate the \profuzz{}'s ability to change the scope of verification by changing the CSR selection.
\texttt{fp-csr} detected the bugs in the floating-point unit (issues \#492, \#493 and \#503) x2.08 times faster compared to the \texttt{selected} configuration while showing a slowdown in detecting other bugs.

We also show the effect of CSR selection on TTE of the bugs through \texttt{all-csr} configuration. 
\texttt{all-csr} configuration failed to detect two of the bugs within the allocated fuzzing time.
Moreover, \texttt{all-csr} is significantly slower (i.e., 0.06$\times$ on average) than \texttt{selected} in detecting bugs.

\begin{table*}[]
\small
\centering
\caption{The speedup achieved by \texttt{selected} \profuzz{} configuration  over no-cov-difuzzrtl, and reg-cov-difuzzrtl for the ground-truth bugs in the BOOM processor. 
We also report speedup of \texttt{fp-csr} and \texttt{all-csr} \profuzz{} configurations over \texttt{selected} \profuzz{} configuration.
In the table, we state the maximum allowed runtime of 48 hours (172800 seconds) for bugs that could not be found.}
\label{tab:tab1}
%\resizebox{\columnwidth}{!}{%
\begin{tabular}{|c|r|rr|rrr|rr|rr|}
\hline
 &
  \multicolumn{1}{c|}{{\begin{tabular}[c]{@{}c@{}}\textbf{no-cov-}\\ \textbf{difuzzrtl}\end{tabular}}} &
  \multicolumn{2}{c|}{{\begin{tabular}[c]{@{}c@{}}\textbf{reg-cov-}\\ \textbf{difuzzrtl}\end{tabular}}}&
  \multicolumn{3}{c|}{{\begin{tabular}[c]{@{}c@{}}\textbf{\profuzz{}}\\ \textbf{(selected)}\end{tabular}}} &
  \multicolumn{2}{c|}{{\begin{tabular}[c]{@{}c@{}}\textbf{\profuzz{}}\\ \textbf{(fp-csr)}\end{tabular}}} & 
  \multicolumn{2}{c|}{{\begin{tabular}[c]{@{}c@{}}\textbf{\profuzz{}}\\ \textbf{(all-csr)}\end{tabular}}} \\ \hline
 {\begin{tabular}[c]{@{}c@{}}Issue\\ No\end{tabular}} &
  \multicolumn{1}{c|}{Time (s)} &
  \multicolumn{1}{c|}{Time (s)} &
  \multicolumn{1}{c|}{\begin{tabular}[c]{@{}c@{}}Speedup\\ (over \\ \texttt{no-cov})\end{tabular}} &
  \multicolumn{1}{c|}{Time (s)} &
  \multicolumn{1}{c|}{\begin{tabular}[c]{@{}c@{}}Speedup\\ (over\\ \texttt{no-cov})\end{tabular}} &
  \multicolumn{1}{c|}{\begin{tabular}[c]{@{}c@{}}Speedup\\ (over\\ \texttt{reg-cov})\end{tabular}} &
  \multicolumn{1}{c|}{Time (s)} &
  \multicolumn{1}{c|}{\begin{tabular}[c]{@{}c@{}}Speedup\\ (over\\ \texttt{selected})\end{tabular}} &
  \multicolumn{1}{c|}{Time (s)} &
  \multicolumn{1}{c|}{\begin{tabular}[c]{@{}c@{}}Speedup\\ (over\\ \texttt{selected})\end{tabular}} \\ \hline
  \hline
\#458 & 104.3    & \multicolumn{1}{r|}{70.3}    & 1.48 & \multicolumn{1}{r|}{54}     & \multicolumn{1}{r|}{1.93} & \multicolumn{1}{r|}{1.3} & \multicolumn{1}{r|}{151324.8} & \multicolumn{1}{r|}{0.0} & \multicolumn{1}{r|}{172800} & \multicolumn{1}{r|}{NA}\\ \hline
\#454 & 32883.3  & \multicolumn{1}{r|}{45322} & 0.73 & \multicolumn{1}{r|}{25020}  & \multicolumn{1}{r|}{1.31} & 1.81 & \multicolumn{1}{r|}{119886.2} & \multicolumn{1}{r|}{0.2} & \multicolumn{1}{r|}{39523.3} & \multicolumn{1}{r|}{0.63}\\ \hline
\#492 & 2047.2   & \multicolumn{1}{r|}{4238.9}  & 0.48 & \multicolumn{1}{r|}{1821.2}   & \multicolumn{1}{r|}{1.12} & 2.32 & \multicolumn{1}{r|}{1221.8} & \multicolumn{1}{r|}{1.49} & \multicolumn{1}{r|}{172800} & \multicolumn{1}{r|}{NA}\\ \hline
\#493 & 585.4    & \multicolumn{1}{r|}{494.9}   & 1.18 & \multicolumn{1}{r|}{278.7}    & \multicolumn{1}{r|}{2.1} & 1.77 & \multicolumn{1}{r|}{170.1} & \multicolumn{1}{r|}{1.63} & \multicolumn{1}{r|}{526.6} & \multicolumn{1}{r|}{0.52} \\ \hline
\#503 & 1463.7   & \multicolumn{1}{r|}{1011.1}  & 1.44 & \multicolumn{1}{r|}{2795.9}   & \multicolumn{1}{r|}{0.52} & 0.36 & \multicolumn{1}{r|}{757.6} & \multicolumn{1}{r|}{3.69} & \multicolumn{1}{r|}{62246.8} & \multicolumn{1}{r|}{0.04} \\ \hline
\#504 & 172800 & \multicolumn{1}{r|}{172800} & NA & \multicolumn{1}{r|}{172800} & \multicolumn{1}{r|}{NA} & NA & \multicolumn{1}{r|}{172800} & \multicolumn{1}{r|}{NA} & \multicolumn{1}{r|}{172800} & \multicolumn{1}{r|}{NA}\\ \hline
\hline
Geo. &
  3182.9 &
  \multicolumn{1}{r|}{3225.9} &
  0.98 &
  \multicolumn{1}{r|}{2630.7} &
  \multicolumn{1}{r|}{1.21} &
  1.23 &
  \multicolumn{1}{r|}{8890.2} &
  \multicolumn{1}{r|}{0.23} & 
  \multicolumn{1}{r|}{43402.2} &
  \multicolumn{1}{r|}{0.06}\\ \hline
\end{tabular}
%}
\end{table*}

\begin{figure}[t]
        \centering
        \begin{subfigure}[b]{0.75\columnwidth}
            \centering
            \includegraphics[width=\columnwidth]{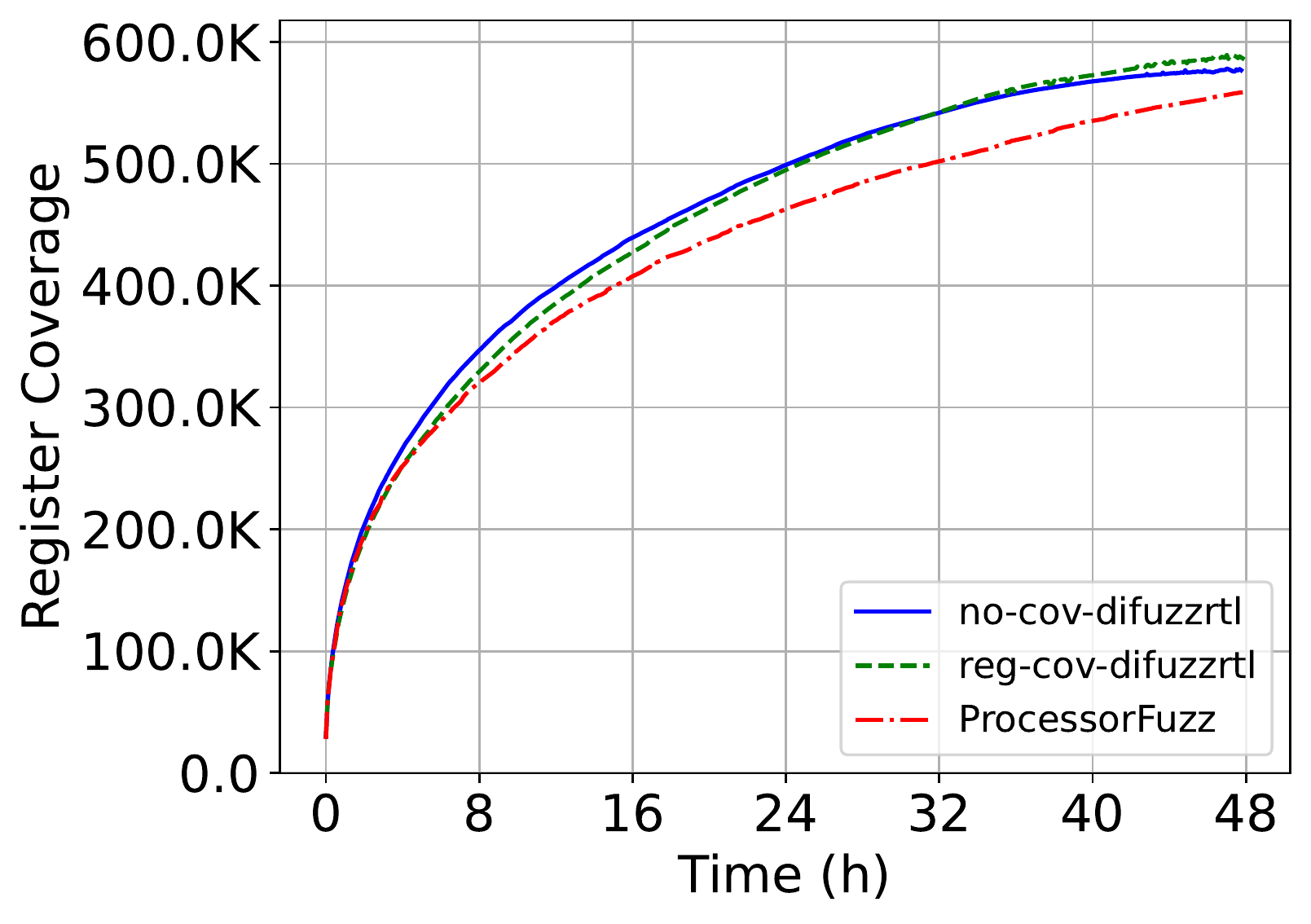}
            \caption[Network2]%
            {{\small Register coverage progress during fuzzing.}}    
            \label{fig:ground-truth-cov}
        \end{subfigure}
        \vfill
        \begin{subfigure}[b]{0.75\columnwidth}  
            \centering 
            \includegraphics[width=\columnwidth]{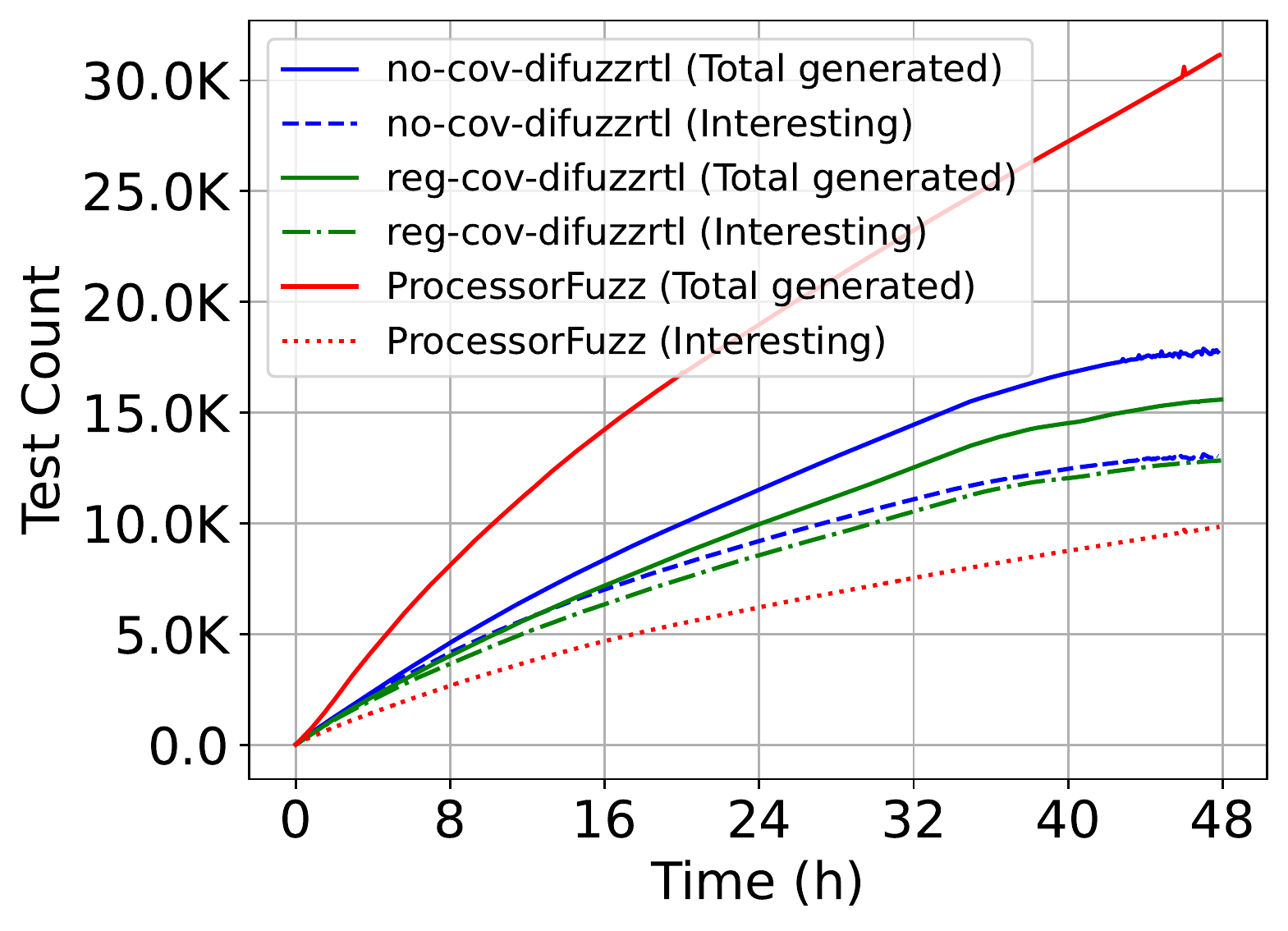}
            \caption[]%
            {{\small Coverage increasing and total test input counts during fuzzing.}}    
            \label{fig:test-num-BOOM}
        \end{subfigure}
        \caption[ Coverage details for different settings ]
        {Coverage details for different settings.} 
        \label{fig:coverage}
    \end{figure}
    
To understand the performance of \profuzz{} and DIFUZZRTL for different bugs, we further study the relationship among register coverage, CSR-transition coverage, and bug-finding times. 
Specifically, in Figure~\ref{fig:ground-truth-cov}, we show the measured register coverage progress for different settings of DIFUZZRTL and \profuzz{}.
Although \profuzz{} covers less number of states (i.e., achieves lower register coverage) during fuzzing, it was still able to discover bugs faster. 
For instance, \profuzz{} triggered the most challenging bug based on the TTE  (i.e., \#454) after exploring 303K states while \texttt{no-cov-difuzzrtl} and \texttt{reg-cov-difuzzrtl} triggered that bug after exploring 364K and 354K states, respectively. 
This particular bug shows that higher register state coverage does not necessarily translate to a faster bug discovery. 
Indeed, an increase in coverage due to value changes in datapath registers can mislead the fuzzer since inputs with similar characteristics (see the multiplication example in Section~\ref{difuz-motiv}) are repeatedly used by the fuzzer to generate a new set of inputs.

In Figure~\ref{fig:test-num-BOOM}, we also show the total number of test inputs that lead to a coverage increase, i.e. `interesting test inputs', and the total number of inputs generated by the mutation engine for the two settings of DIFUZZRTL and \profuzz{}.
For \texttt{no-cov-difuzzrtl} and \texttt{reg-cov-difuzzrtl}, we use the register coverage metric, same as that used in the DIFUZZRTL work, to realize if a test input increases coverage.
For \profuzz{}, we use the CSR-transition coverage metric to detect inputs that resulted in a coverage increase.
The results provide an important takeaway. 
Although \profuzz{} generates significantly more inputs than other approaches, it is very selective when categorizing a test input as an 'interesting' input. 
Consequently, \profuzz{} identified only 33\% of the generated test inputs as interesting (i.e., caused a unique CSR transition).
Moreover, \profuzz{} could expose the bugs faster although it used the least number of test inputs for RTL simulation.
Note that \profuzz{} launched the RTL simulation only with interesting inputs (i.e., curved dotted red line) and discarded any other generated input.
Using the fast ISA simulation enabled \profuzz{} to quickly eliminate inputs that do not result in a new FSM state and spend more time on inputs that explore new FSM states.

\begin{figure}[]
        \centering
        \begin{subfigure}[b]{0.48\columnwidth}
            \centering
            \includegraphics[width=\columnwidth]{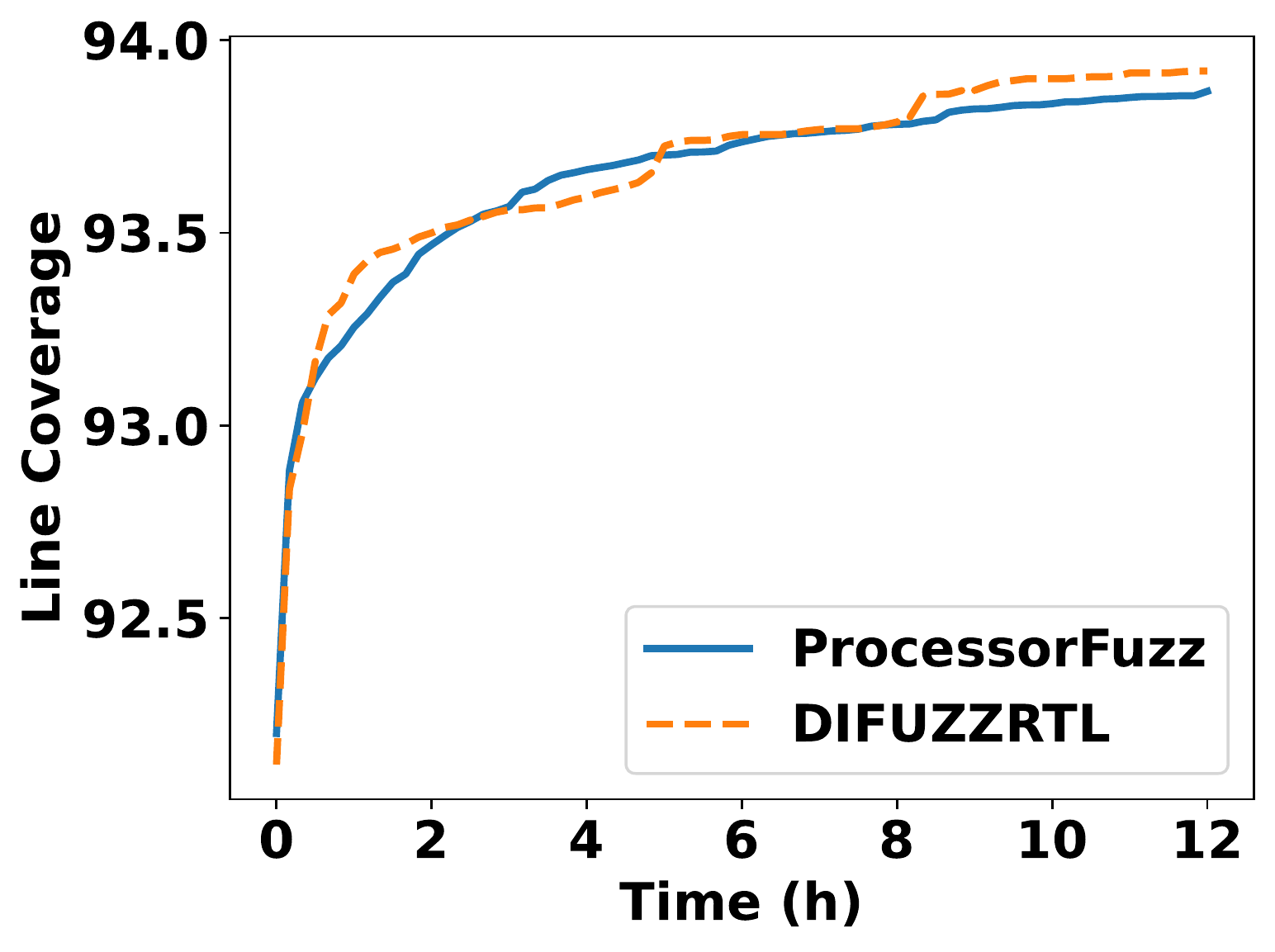}
            %\caption[Network2]%
            %{{\small Line coverage progress during fuzzing.}}    
            \label{fig:line-cov}
        \end{subfigure}
        \begin{subfigure}[b]{0.48\columnwidth}
            \centering
            \includegraphics[width=\columnwidth]{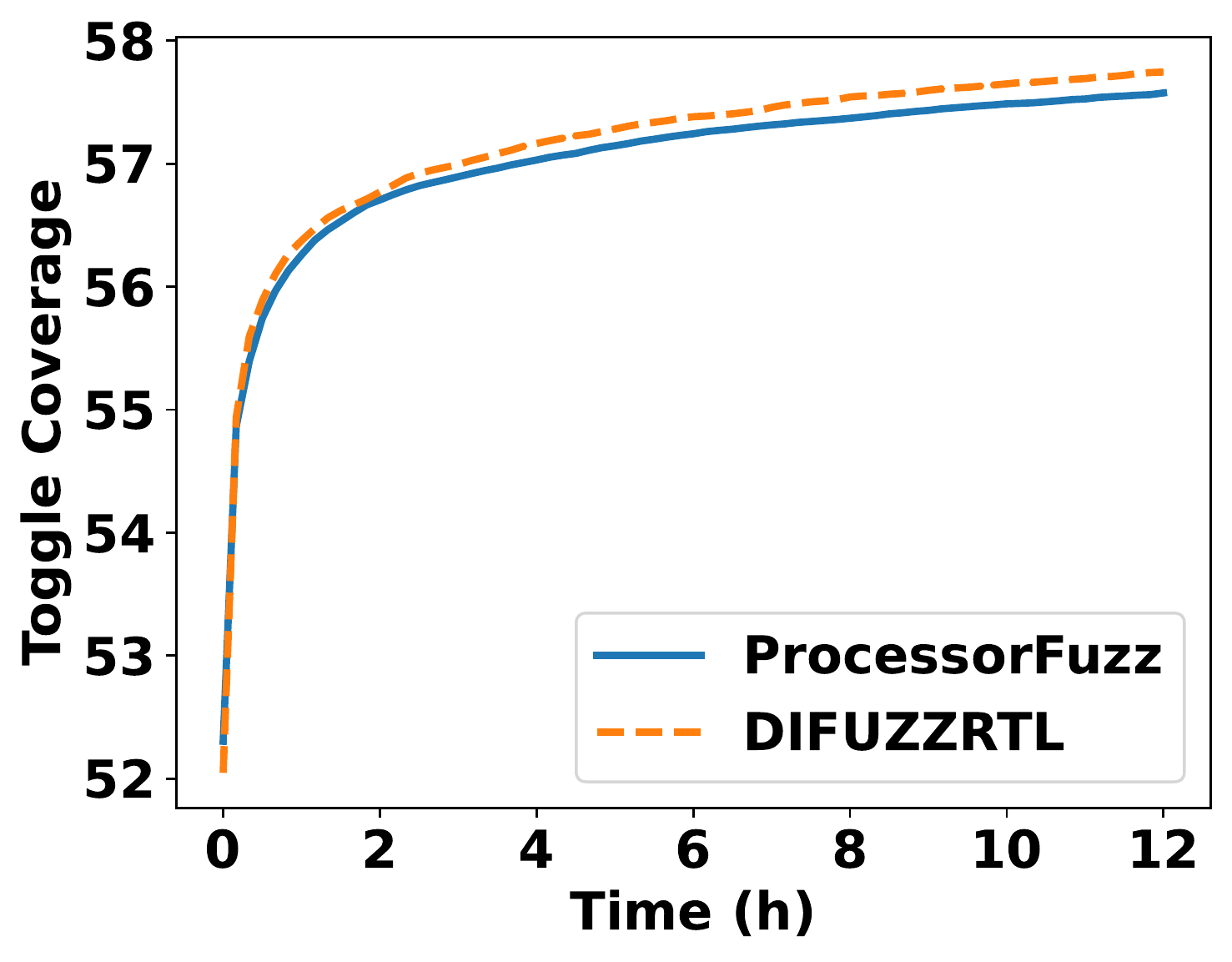}
            %\caption[Network2]%
            %{{\small Toggle coverage progress during fuzzing.}}    
            \label{fig:toggle-cov}
        \end{subfigure}
        \vfill
        \begin{subfigure}[b]{0.48\columnwidth}  
            \centering 
            \includegraphics[width=\columnwidth]{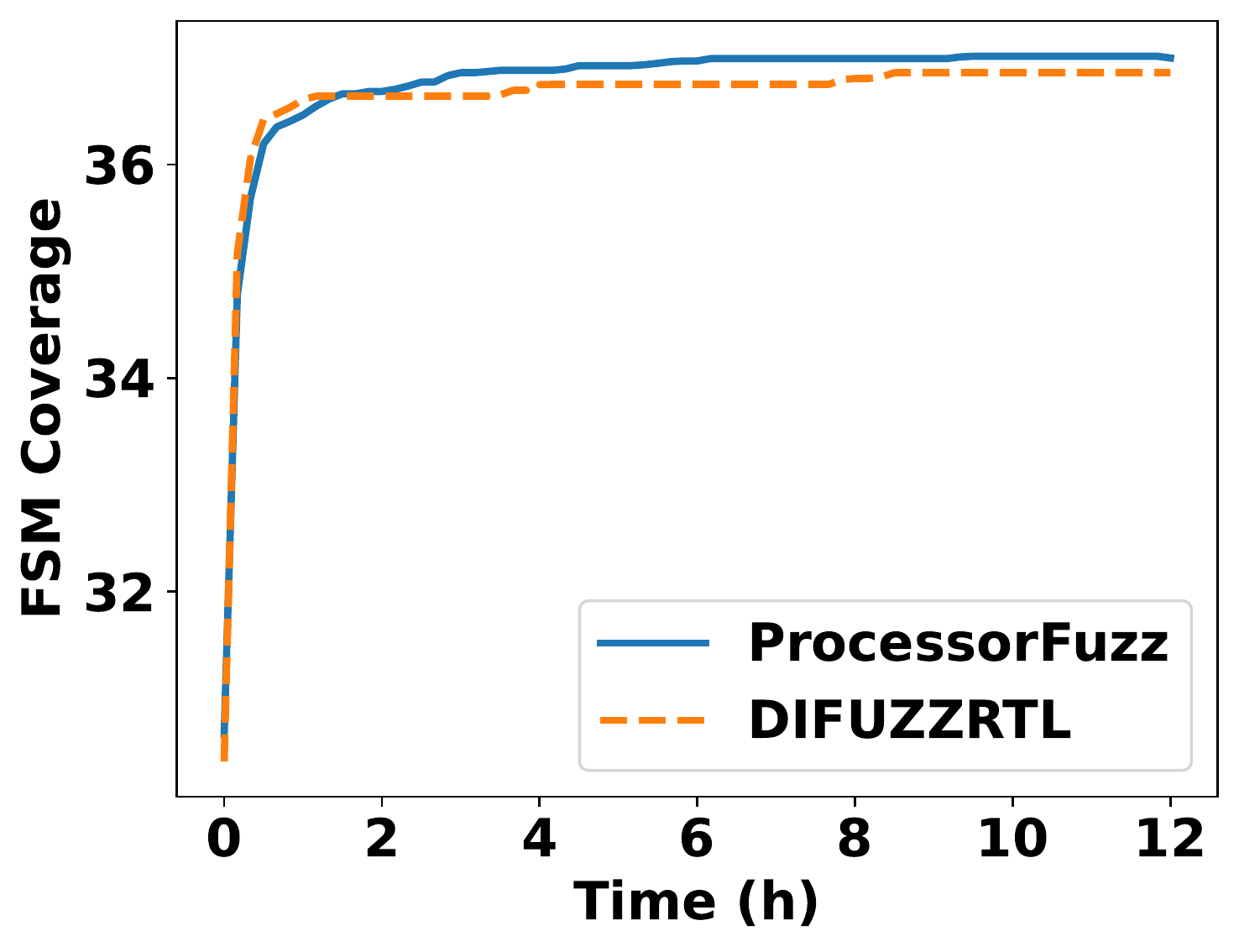}
            %\caption[]%
            %{{\small FSM coverage progress during fuzzing.}}    
            \label{fig:fsm-cov}
        \end{subfigure}
        \begin{subfigure}[b]{0.48\columnwidth}  
            \centering 
            \includegraphics[width=\columnwidth]{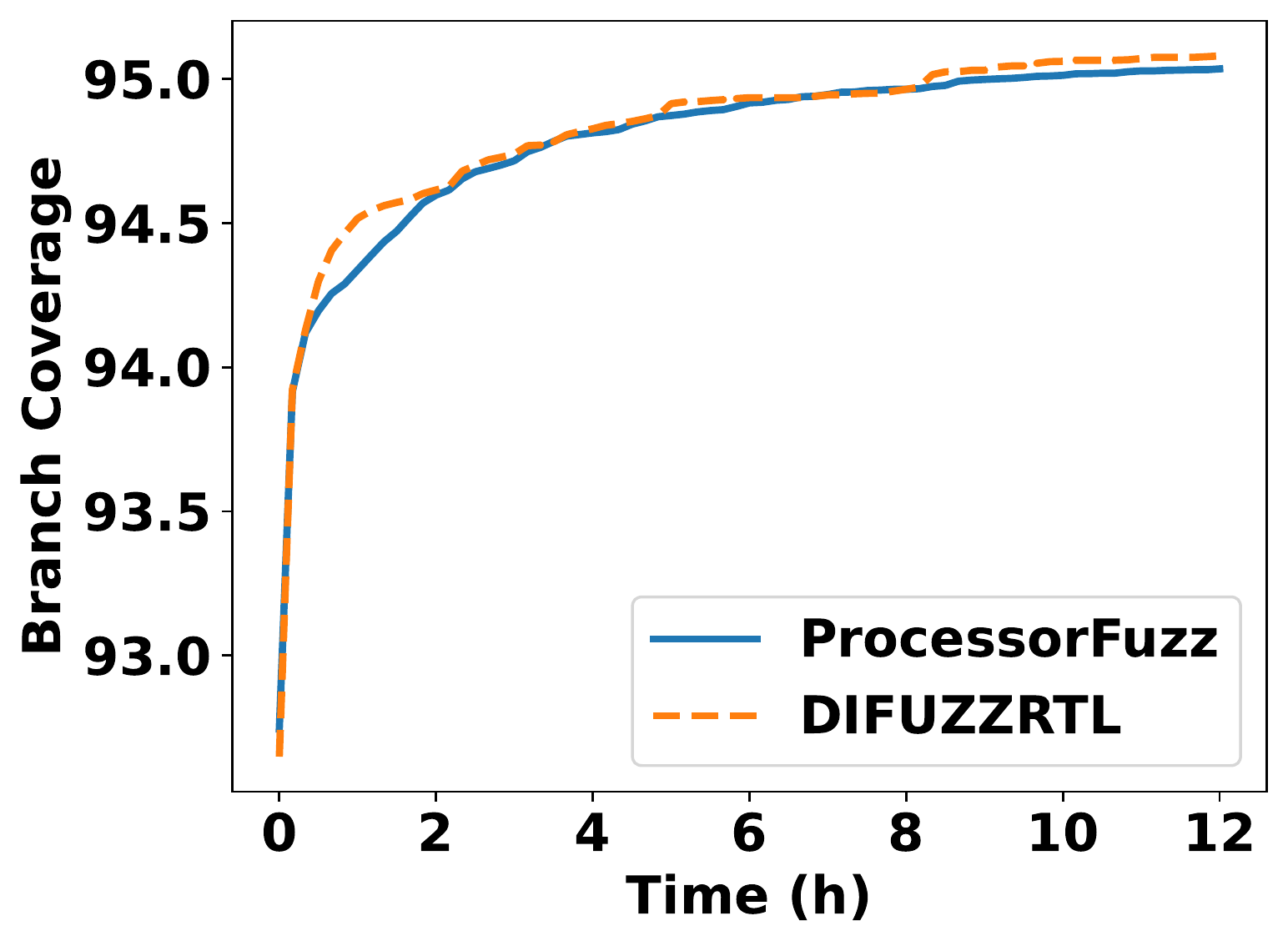}
            %\caption[]%
            %{{\small Branch coverage progress during fuzzing.}}    
            \label{fig:fsm-cov}
        \end{subfigure}
        \caption[ Progress of standard RTL coverage metrics for BOOM core during fuzzing ]
        {The progress of industry-standard RTL coverage metrics for \profuzz{} and \difuz{} for BOOM core. } 
        \vspace{-0.5cm}
        \label{fig:standard_coverage}
    \end{figure}

In Figure \ref{fig:standard_coverage}, we show how ProcessorFuzz performs in terms of industry standard RTL coverage metrics (i.e., line, toggle, FSM, and branch coverage) for BOOM core.
Here, our goal is to present the effectiveness of \profuzz{} based on the widely-used coverage metrics.
In particular, we aim to explore how well the CSR-transition coverage metric is able to result in test cases that cover different lines, toggles, FSMs, and branches in the processor RTL. 
We compare \profuzz{}'s overall coverage based on these four metrics against \difuz{}. 
Line coverage represents the percentage of RTL code lines that got exercised during the simulation. 
Toggle coverage indicates the percentage of bits in wires and registers that toggled during the simulation. 
FSM coverage represents the percentage of FSM states reached during the simulation. 
Branch coverage represents the percentage of different branches that was taken during the simulation against the total branches in the design.  
In particular, we obtained all the seeds generated by each approach during fuzzing and feed them to the Synopsys VCS tool one by one and reported coverage.
Note that Synopsys VCS tool is significantly slower when collecting coverage, and therefore, we could report coverage progress for the first 12 hours of fuzzing although we run VCS tool for a week. 
The main takeaway from Figure \ref{fig:standard_coverage} is that, even though \profuzz{} uses CSR-transition coverage extracted from ISA simulation, \profuzz{} performs as well as \difuz{} in terms of standard RTL coverage metrics and is able to cover different RTL regions based on different metrics. 

\subsection{Newly Discovered Bugs}

\begin{table*}[]
\small
\centering
\caption{Brief description of bugs discovered by \profuzz{}, and their current status, in various processor cores.
% {\color{red}Make sure the wording used for describing the bug does not match with the description used when filing the bug. Otherwise our identity will get revealed and the paper will get rejected.}
}
\label{tab:bugDescription}
\begin{tabular}{|c|l|p{10cm}|l|}
\hline
Bug &
  \multicolumn{1}{c|}{\begin{tabular}[c]{@{}c@{}}Core / \\ Simulator\end{tabular}} &
  \multicolumn{1}{c|}{Brief description of the bug} &
  \multicolumn{1}{c|}{Status} \\ \hline
1 &
  BlackParrot &
  Non-boxed single-precision floating point values are not interpreted as NaNs & 
  %ORIGINAL: Some FPU operations not producing NaNs as they should
 Confirmed; not fixed\\ \hline

2 &
  BlackParrot &
  \begin{tabular}[c]{@{}l@{}}Read-after-Write dependencies on \texttt{fcsr.fflags} are not satisfied.\end{tabular} &
  %ORIGINAL: fcsr.fflags RAW hazards not detected
  \begin{tabular}[c]{@{}l@{}}Fixed\end{tabular} \\ \hline
3 &
  BlackParrot &
  When \texttt{mstatus.FS} is not set and the \texttt{fcsr} is written, \texttt{FS} is unexpectedly updated.&
  %ORIGINAL: BP sets FS field to DIRTY during exception when FS is OFF
  \begin{tabular}[c]{@{}l@{}}Fixed\end{tabular} \\ \hline
4 &
  BlackParrot &
  \begin{tabular}[c]{@{}l@{}} The 2 low-bits of \texttt{sepc} CSR are not write-insensitive.\end{tabular} &
  %ORIGINAL: BP wrongly sets sepc[0]
  \begin{tabular}[c]{@{}l@{}}Fixed\end{tabular} \\ \hline
5 &
  BlackParrot &
  \begin{tabular}[c]{@{}l@{}}No exception raised when writing certain read-only CSRs.\end{tabular} &
  %ORIGINAL: BP does not raise exception when attempting to update mhartid
  \begin{tabular}[c]{@{}l@{}}Fixed\end{tabular} \\ \hline

6 &
  BlackParrot &
  Reading \texttt{zero} register, following specific instruction sequences, return unexpected non-zero values & 
  %ORIGINAL: Division by zero results are incorrect according to the specification
%\rewrite{ When the destination register of a division instruction is \texttt{x0}, BP reads \texttt{x0} as non-zero value, which causes an issue in case the next instruction reads \texttt{x0}.} &
     \begin{tabular}[c]{@{}l@{}}Fixed\end{tabular} \\ \hline
7 &
  BlackParrot &
  \begin{tabular}[c]{@{}l@{}} Unexpected store access-fault on properly aligned, unpaired \textit{sc.d} instruction. \end{tabular} &
  Reported\\ \hline 
8 &
  Dromajo &
  PMP checks are performed, and raise exceptions upon encountering violations, even with no PMP entries set.&
  %ORIGINAL: Pagetable walk throws access violation even with no PMP entries set -- not sure 
  Confirmed; not fixed.\\ \hline
9 &
  \begin{tabular}[c]{@{}c@{}}Rocket \& \\ BOOM\end{tabular} &
  Instruction page fault not raised when accessing non-leaf PTEs with certain unspecified page attributes. &
  %ORIGINAL: Instruction page fault when D/A/U bits set in a non-leaf PTE
  Fixed \\ \hline
10 &
  BOOM &
  \texttt{mstatus.FS} is gratuitously set to dirty. &
  %ORIGINAL: FS is not set to dirty when mstatus is written
 Confirmed; not fixed\\ \hline

 \end{tabular}
\end{table*}
In Table~\ref{tab:bugDescription}, we document the various \textbf{new} bugs discovered by \profuzz{} in the selected processors mentioned earlier and in the ISA simulator used as a reference model. 
In the following subsections, we describe and highlight the significance of each bug.

\subsubsection{Bug Descriptions}

\noindent\textbf{Bug 1.} When multiple floating-point precisions are supported by a floating-point unit in a processor, valid lower precision values are expected to be NaN-boxed (i.e., remaining upper bits set to 1's).  
Otherwise, lower precision values are expected to be interpreted as NaNs.
BlackParrot does not interpret non-boxed floats as NaNs, which leads to functionally incorrect computations on non-boxed floats, thus violating the ISA specification.
Incorrect computations in security-critical functions (e.g., in cryptographic applications)  can compromise the security of a processor (CWE-1201~\cite{cwe}).

\noindent\textbf{Bug 2.} Certain floating-point instructions update \texttt{fflags} CSR which holds events like floating-point overflow, division by zero, etc. 
A read-after-write (RAW) hazard occurs in a pipelined processor when a floating-point instruction that writes to \texttt{fflags} is followed by an instruction that reads \texttt{fflags}.%When certain floating-point instructions (e.g., \texttt{fdiv}) are in the pipeline of a processor, there is a read-after-write (RAW) dependency established on the \texttt{fflags} register which holds events like floating-point overflow, division by zero, etc.
\profuzz{} detected that this particular hazard is not handled by the BlackParrot processor, causing the software to read an outdated \texttt{fflags} value.%A pipelined processor must implement the necessary logic to handle the hazard. Otherwise, software can read an outdated \texttt{fflags} value.
The RISC-V ISA requires explicit checks of the \texttt{fflags} CSR in software to identify floating-point overflows, invalid operations, etc.
Therefore, failure to detect an overflow can lead to security issues in software such as buffer overflows (e.g., CVE-2020-10029~\cite{fpsve}).
This bug falls under core and compute hardware weakness (CWE-1201~\cite{cwe}). 
\profuzz{} was able to detect this bug since it monitors the transitions in \texttt{fflags}.

\noindent\textbf{Bug 3.} When the \texttt{FS} field of \texttt{mstatus} CSR is set to 0, it indicates that floating-point extension is turned off. 
In such cases, accessing \texttt{fcsr} is expected to raise an illegal instruction exception without altering the \texttt{FS} field of \texttt{mstatus} CSR.
However, BlackParrot wrongly sets the \texttt{FS} field to \textit{dirty}, instead of keeping it unchanged. 

\noindent\textbf{Bug 4.} The least significant two bits of \texttt{sepc} CSR must be always hardwired to 0 on implementations that only support 32-bit instruction alignment. 
Any write from software to the least significant two bits of the \texttt{sepc} CSR must be discarded. 
However, BlackParrot updated the low bits when a test input attempted to modify them. 
Further analysis showed that this issue exists for \texttt{mepc} CSR as well. 

\noindent\textbf{Bug 5.} Any attempt to modify a read-only register is supposed to trap with an illegal instruction exception. 
\profuzz{} discovered that BlackParrot does not raise an illegal instruction exception if a test input updates a read-only register, specifically \texttt{mhartid}.
\profuzz{} monitors \texttt{mcause} and \texttt{scause} CSRs that are in charge of exception handling and was, therefore, able to expose this bug.

\noindent\textbf{Bug 6.} Any write attempt to the zero register (i.e., \texttt{x0}) must be ignored according to the RISC-V ISA.
However, in the BlackParrot processor, we detected that the \texttt{x0} register is read as a non-zero value one of the preceding division instructions that writes to \texttt{x0} is still in the pipeline.
Further analysis revealed that this discrepancy is due to bypassing the result of division operation to the following instruction even when the destination register of a division operation is \texttt{x0}. 
\profuzz{} was able to identify this bug because a test input that has this scenario resulted in a CSR transition in \texttt{fflags} due to division by zero.

\noindent\textbf{Bug 7.}
A store-conditional instruction, if properly aligned to the appropriate word boundary, should not raise a store-access fault. 
However, BlackParrot raises a store access fault when executing an unpaired, but properly aligned \textit{sc.d} instruction. 
We reported the issue to the BlackParrot designers and are currently waiting for their response.

\noindent\textbf{Bug 8.}
According to RISC-V privileged specification, the effective privilege mode for implicit page table accesses should be supervisor mode.
However, we observed that Dromajo accesses page tables in user mode privilege level when executing user-mode programs. 
Further analysis revealed that Dromajo also carries out Physical Memory Protection (PMP) checks in user mode when no PMP entries are set, violating the RISC-V ISA privileged specification in two counts.

\noindent\textbf{Bug 9.} In a multi-level page table implementation, the accessed (A), dirty (D), and user-mode (U) bits of a non-leaf page table entry (PTE) are reserved for future use and should be cleared. 
If these bits are set in a non-leaf PTE, the processor must raise an instruction page fault when accessing the PTE according to RISC-V ISA. 
We discovered that Rocket and BOOM cores do not raise instruction page fault when software attempts to access a PTE with any of A, D, or U bits set. 
This bug is similar to CWE-1209\cite{cwe} where failure to disable reserved bits allows attackers to compromise the hardware state. 

\noindent\textbf{Bug 10.}
The \texttt{FS} field in the \texttt{mstatus} CSR in RISC-V ISA is used to check whether save and restore of floating-point registers are required when there is a context switch.
\profuzz{} detected that BOOM set the \texttt{FS} field to dirty for any write to fcsr register, even when the value of fcsr is zero and unchanged by the write operation.
This scenario is not a violation of RISC-V ISA due to the flexibility allowed by the ISA for maintaining \texttt{FS} field. 
Nevertheless, setting \texttt{FS} field when the floating-point unit state is unchanged degrades the performance as the processor unnecessarily saves and restores floating-point registers.

% \href{https://github.com/black-parrot/black-parrot/issues/971}{\#971} & 464.9    & 230.2    & 2.0 \\ \hline
% \href{https://github.com/chipsalliance/dromajo/issues/50}{\#50}       & 5192.7   & 3264.6   & 1.6 \\ \hline
% \href{https://github.com/black-parrot/black-parrot/issues/994}{\#994} & 95695.0  & 57441.3  & 1.7 \\ \hline
% \href{https://github.com/black-parrot/black-parrot/issues/969}{\#969} & 1520.1   & 1474.5   & 1.0 \\ \hline
% \href{https://github.com/black-parrot/black-parrot/issues/970}{\#970} & 585.3    & 308..0   & 1.9 \\ \hline
% \href{https://github.com/black-parrot/black-parrot/issues/967}{\#967} & 476.1    & 242.1    & 2.0 \\ \hline
% \href{https://github.com/black-parrot/black-parrot/issues/832}{\#832} & 172800.0 & 147942.3 & 1.2 \\ \hline
% 8 & 132036.1 & 137863.1 & 1.0 \\ \hline

% \begin{table}[!t]
% \small
% \centering
% \caption{The speedup achieved by \profuzz{} on newly discovered bugs in the BlackParrot processor. A maximum of 48 hours (172800 seconds) is assigned if a bug is not triggered.}
% \label{tab:tab2}

% \begin{tabular}{|c|r|r|r|}
% \hline
% \textbf{Bug} &
%   \multicolumn{1}{c|}{\textbf{no-cov-difuzzrtl}} &
%   \multicolumn{1}{c|}{\textbf{\profuzz{}}} &
%   \multicolumn{1}{c|}{\textbf{Speedup}} \\ \hline
%   \hline
% %% Prior to reordering blackparrot bugs to be placed together
% Bug 1 & 464.9    & 230.2    & 2.0 \\ \hline
% Bug 2 & 95695.0  & 57441.3  & 1.7 \\ \hline
% Bug 3 & 1520.1   & 1474.5   & 1.0 \\ \hline
% Bug 4 & 585.3    & 308.0   & 1.9 \\ \hline
% Bug 5 & 476.1    & 242.1    & 2.0 \\ \hline
% Bug 6 & 172800.0 & 147942.3 & 1.2 \\ \hline
% Bug 7 & 5192.6 & 3264.6 & 1.6 \\ \hline
% \hline
% Geo. &
%   6217.3 &
%   4199.83 &
%   1.6 \\ \hline
  
% \end{tabular}

% \end{table}

\begin{table*}[]
\small
\centering
\caption{The speedup achieved by \profuzz{} over no-cov-difuzzrtl, and reg-cov-difuzzrtl for the ground-truth bugs in the BOOM processor. 
We also report speedup of \texttt{fp-csr} and \texttt{all-csr} \profuzz{} configurations over \texttt{selected} \profuzz{} configuration.
In the table, we state the maximum allowed runtime of 48 hours (172800 seconds) for bugs that could not be found.}
\label{tab:tab2}
\begin{tabular}{|c|r|rr|rrr|rr|rr|}
\hline
 &
  \multicolumn{1}{c|}{\textbf{no-cov-difuzzrtl}} &
  \multicolumn{2}{c|}{\textbf{\profuzz{} (selected)}} &
  \multicolumn{3}{c|}{\textbf{\profuzz{} (\texttt{all-csr})}} &
  \multicolumn{3}{c|}{\textbf{\profuzz{} (\texttt{fp-csr})}}  \\ \hline
 Bug &
  \multicolumn{1}{c|}{Time (s)} &
  \multicolumn{1}{c|}{Time (s)} &
  \multicolumn{1}{c|}{\begin{tabular}[c]{@{}c@{}}Speedup\\ (over \texttt{no-cov})\end{tabular}} &
  \multicolumn{1}{c|}{Time (s)} &
  \multicolumn{1}{c|}{\begin{tabular}[c]{@{}c@{}}Speedup\\ (over \texttt{no-cov})\end{tabular}} &
  \multicolumn{1}{c|}{\begin{tabular}[c]{@{}c@{}}Speedup\\ (over \texttt{selected})\end{tabular}} &
  \multicolumn{1}{c|}{Time (s)} &
  \multicolumn{1}{c|}{\begin{tabular}[c]{@{}c@{}}Speedup\\ (over \texttt{no-cov})\end{tabular}} &
    \multicolumn{1}{c|}{\begin{tabular}[c]{@{}c@{}}Speedup\\ (over \texttt{selected})\end{tabular}} \\
  \hline
 1 & 464.9    & \multicolumn{1}{r|}{230.2}    & 2.02 & \multicolumn{1}{r|}{430.2}     & \multicolumn{1}{r|}{1.08} & \multicolumn{1}{r|}{0.54} & \multicolumn{1}{r|}{1608.7} & \multicolumn{1}{r|}{0.29} & \multicolumn{1}{r|}{0.14} \\ \hline
 2 & 95695.0  & \multicolumn{1}{r|}{57441.3} & 1.67 & \multicolumn{1}{r|}{100804.9}  & \multicolumn{1}{r|}{0.95} & 0.57 & \multicolumn{1}{r|}{122076.0} & \multicolumn{1}{r|}{0.78} & \multicolumn{1}{r|}{0.47} \\ \hline
 3 & 1520.1   & \multicolumn{1}{r|}{1474.5}  & 1.03 & \multicolumn{1}{r|}{921.8}   & \multicolumn{1}{r|}{1.65} & 1.60 & \multicolumn{1}{r|}{172800} & \multicolumn{1}{r|}{0.01} & \multicolumn{1}{r|}{0.01} \\ \hline
 4 & 585.3    & \multicolumn{1}{r|}{308.0}   & 1.90 & \multicolumn{1}{r|}{558.8}    & \multicolumn{1}{r|}{1.05} & 0.55 & \multicolumn{1}{r|}{13560.4} & \multicolumn{1}{r|}{0.04} & \multicolumn{1}{r|}{0.02}  \\ \hline
 5 & 476.1   & \multicolumn{1}{r|}{242.1}  & 1.97 & \multicolumn{1}{r|}{239.7}   & \multicolumn{1}{r|}{1.99} & 1.01 & \multicolumn{1}{r|}{39150.9} & \multicolumn{1}{r|}{0.01} & \multicolumn{1}{r|}{0.01}  \\ \hline
 6 & 172800 & \multicolumn{1}{r|}{147942.3} & 1.17 & \multicolumn{1}{r|}{148655.0} & \multicolumn{1}{r|}{1.16} & 1.00 & \multicolumn{1}{r|}{172800} & \multicolumn{1}{r|}{1.00} & \multicolumn{1}{r|}{0.86} \\ \hline
 7 & 5192.6 & \multicolumn{1}{r|}{3264.6} & 1.59 & \multicolumn{1}{r|}{172800} & \multicolumn{1}{r|}{0.03} & 0.02 & \multicolumn{1}{r|}{172800} & \multicolumn{1}{r|}{0.03} & \multicolumn{1}{r|}{0.02} \\ \hline
\hline
Geo. &
  4018.1&
  \multicolumn{1}{r|}{2550.5} &
  1.58 &
  \multicolumn{1}{r|}{5420.9} &
  \multicolumn{1}{r|}{0.74} &
  0.47 &
  \multicolumn{1}{r|}{47404.8} &
  \multicolumn{1}{r|}{0.08} & 
  \multicolumn{1}{r|}{0.05} \\ \hline
\end{tabular}
%}
\end{table*}

\subsubsection{Timing Results}
In Table~\ref{tab:tab2}, we provide the TTEs for six newly identified and confirmed bugs (Bug 1-6) and one newly identified but currently waiting confirmation bug (Bug 7) in BlackParrot core. 
We did not include Bug 8-10 since they were easily detected in all the settings that we used in our evaluation.
For this evaluation, we were only able to compare \profuzz{} with \texttt{no-cov-difuzzrtl}.
As detailed in Section~\ref{sec:settings}, we could not instrument BlackParrot with register coverage since \difuz{} lacks support for SystemVerilog (detailed in Section~\ref{sec:settings}).
\profuzz{} does not require any instrumentation on the RTL design, therefore, could successfully guide the fuzzer with CSR-transition coverage to expose bugs.
Overall, \profuzz{} achieved 1.6$\times$ speedup, on average, over \texttt{no-cov-difuzzrtl}.
Note that only \profuzz{} was able to detect Bug 6 from Table~\ref{tab:bugDescription}.
Similar to the experiment that we conducted in the BOOM processor using the ground-truth bugs, \texttt{all-csr} configuration of \profuzz{} performed poorly compared to \texttt{selected} configuration (i.e., 0.47$\times$ slow-down).
Moreover, \texttt{fp-csr} configuration identified floating-point related bugs fairly faster (e.g., Bug 2) compared to other type of bugs (e.g., Bug 4 that focuses on \texttt{sepc} CSR).
\section{ Related Work }
\label{sec:relatedwork}
We divide the related work into three different categories. 
First, we present traditional methods in hardware verification such as random instruction generation, coverage-directed test generation, and formal verification.
Then, we present fuzzing-based hardware verification approaches and how \profuzz{} differs from existing fuzzing works.
Finally, we discuss the usage of differential testing in the software domain. 

\subsection{Traditional Hardware Verification}
\label{subsec:trad-hw-ver}
Random instruction generators~\cite{riscv-torture, riscvdv, force-riscv,aapg,herdt2020efficient} have been commonly used in processor verification since they require limited human expertise and are scalable to large RTL designs.
These tools produce random assembly programs based on a set of constraints such as the instruction mix, frequencies, etc., to identify functional bugs in processors.
The lack of coverage guidance in these tools leads to the generation of the repetitive inputs that test the same processor functionalities, thereby decreasing the chances of finding bugs~\cite{hur2021difuzzrtl,laeufer2018rfuzz}.

A verification engineer can target the uncovered RTL regions by adjusting the constraints that control the random test generator.
For instance, if coverage is maximized in the branch prediction unit but not in the load-store unit, the verification engineer can increase the ratio of load and store instructions.
However, this method significantly increases engineering effort, and therefore, slows down the verification process.
To overcome this problem, researchers proposed several coverage-directed test generation mechanisms~\cite{fine2003coverage, wagner2005stresstest,tasiran2001functional,nativ2001cost,gal2021automatic,bose2001genetic,squillero2005microgp} that automatically direct the next round of test generation that targets the uncovered parts of RTL. 
These works use RTL simulators to dynamically monitor the behavior of an RTL design and adjust the test generator constraints towards producing tests inputs that target uncovered RTL regions.
Unfortunately, these works are generally DUT-specific which hinders their general applicability.

Formal verification methods (e.g., symbolic execution,  model checking) are also widely used in hardware verification\cite{chen2011property,mukherjee2015hardware,JasperGold}. 
These methods use mathematical reasoning to prove that a hardware design conforms to its specification.
Unfortunately, formal verification methods have a well-known state explosion problem, and therefore, do not scale well for complex RTL designs such as a processor~\cite{dessouky2019hardfails}. 
Indeed, a prior work~\cite{dessouky2019hardfails} clearly presented that many processor bugs cannot be identified with these tools due to the space explosion issues and emphasised the necessity of complementary approaches to existing formal verificaiton tools.
\subsection{Hardware Fuzzing} 
\label{subsec:hw-fuzzing}

Over the past few years, fuzzing has gained traction in RTL verification due to its bug-finding success in the software domain~\cite{ossfuzz-bug}.
In Table~\ref{tab:relatedwork}, we provide a high-level overview of all fuzzing-based RTL verification approaches. 
For each approach, we include the input format, the coverage metric used to guide the fuzzer, and the method to identify bugs.  
\begin{table*}[]
%\small
\caption{Existing RTL Fuzzers.}
\label{tab:relatedwork}
\resizebox{\textwidth}{!}{%
\begin{tabular}{|c|c|c|c|c|}
\hline
\multicolumn{1}{|l|}{} & \textbf{Input Format} &\textbf{Coverage Metric} & \textbf{Evaluated RTL Designs}                                                                               & \textbf{Bug Discovery Method} \\ \hline \hline
\textbf{RFUZZ~\cite{laeufer2018rfuzz}}         & A Series of Bits   & Mux Toggle        & \begin{tabular}[c]{@{}c@{}}Peripherals, \\ RISC-V Processors (Sodor 1-3-5)\end{tabular}     & Assertion                 \\ \hline
 \textbf{Li et. al~\cite{li2021symbolic}}       & A Series of Bits  & Full Mux Toggle     & \begin{tabular}[c]{@{}c@{}} Custom RISC-V Processor, OpenCore 1200\end{tabular} &  Assertion             \\ \hline
\textbf{DIFUZZRTL~\cite{hur2021difuzzrtl}}     & Assembly    & Register  Coverage   & \begin{tabular}[c]{@{}c@{}}RISC-V Processors \\ (BOOM, Mork1x, Rocket Chip)\end{tabular}   & Golden Model              \\ \hline
 \textbf{DirectFuzz~\cite{canakci2021directfuzz}}    & A Series of Bits & Mux Toggle          & Same as RFUZZ                                                                              & Assertion \\ \hline
 \textbf{Trippel et al.~\cite{trippel2021fuzzing}}       & 
Byte Sequence & Edge Coverage & \begin{tabular}[c]{@{}c@{}} RISC-V IP Cores \\ (AES, HMAC, KMAC, Timer)\end{tabular} &  \begin{tabular}[c]{@{}c@{}} Golden Model \\ Assertion \end{tabular}             \\ \hline
\textbf{TheHuzz~\cite{tyagi2022thehuzz}}     & Assembly & \begin{tabular}[c]{@{}c@{}}  Branch, Line, Statement,  \\ Expression, DFF Toggle, FSM \end{tabular}    & \begin{tabular}[c]{@{}c@{}}RISC-V Processors ( Rocket Chip, CVA6),\\ mor1kx, OpenCore 1200 \end{tabular}   & Golden Model              \\ \hline
\textbf{HYPERFUZZER~\cite{muduli2020hyperfuzzing}}       & A Series of Bits  & High-Level & Custom SoC  & Property Check   \\ \hline
\textbf{Logic Fuzzer~\cite{kabylkas2021effective}}       & \begin{tabular}[c]{@{}c@{}} A Series of Bits,\\ Random Data\end{tabular}  & N/A    & \begin{tabular}[c]{@{}c@{}} RISC-V Processors \\ (BlackParrot, BOOM, CVA6)\end{tabular} &  Golden Model             \\ \hline
\textbf{\profuzz{} (this work)}       & Assembly   &\begin{tabular}[c]{@{}c@{}} Control Path Register, \\ ISA-Sim Transition\end{tabular}     & \begin{tabular}[c]{@{}c@{}}RISC-V processors\\ (BOOM, BlackParrot, Rocket Chip)\end{tabular} & Golden Model              \\ \hline
\end{tabular}
}
\end{table*}

% \textbf{INTROSPECTRE~\missingcitation{}}       & Assembly  & N/A     & \begin{tabular}[c]{@{}c@{}} BOOM\end{tabular} &  Trace Analysis             \\ \hlinehttps://www.overleaf.com/project/60ec63d0bb54d83ec74a4ded
%% \textbf{MicroGP~\cite{squillero2005microgp}}       & Assembly  & Statement         & 5-stage DLX processor  & Golden Model              \\ \hline

RFUZZ~\cite{laeufer2018rfuzz} defines a simple input format (i.e., a series of bits) to increase the portability of hardware fuzzing to a wide range of RTL designs.
Unfortunately, this input format is not effective when fuzzing processors since a processor requires instructions defined by an ISA.
RFUZZ also proposes a new coverage metric, the multiplexer toggle coverage. 
RFUZZ monitors all the multiplexers in the RTL design. 
It retains an input for further mutations if the input toggles a previously uncovered multiplexer selection signal.
A follow-up work by Li et al.~\cite{li2021symbolic} enhances RFUZZ with symbolic simulation and defines a full multiplexer toggle coverage metric that counts a multiplexer signal as covered for either 1-0-1 or 0-1-0 toggles.
Both RFUZZ and Li et al. are highly coupled to Chisel HDL which limits the applicability of the approach~\cite{sadeghi2021organizing}. 
Additionally, monitoring multiplexers in complex designs introduces excessive performance overhead~\cite{hur2021difuzzrtl}.
\profuzz{} is agnostic to HDL and also does not require any instrumentation in the HDL code, which makes it both practical and efficient.

DIFUZZRTL monitors registers that directly or indirectly control multiplexer selection signals.
This design choice makes it more efficient than RFUZZ since the total number of bits in the identified registers is significantly less than multiplexers. 
Moreover, DIFUZZRTL shows that RFUZZ's coverage metric does not precisely capture the FSM states.
To mitigate this issue, DIFUZZRTL monitors value changes in the identified registers for each cycle.
Unfortunately, DIFUZZRTL monitors many registers in the datapath as well, thereby misguiding the fuzzer as detailed in Section~\ref{difuz-motiv}.

Trippel et al.~\cite{trippel2021fuzzing} translate hardware designs to software models and fuzzes those models.
This way, available coverage metrics used by software fuzzers (e.g., basic block, edge) can be used for fuzzing hardware as well.
However, this method of converting hardware designs to software models introduces additional challenges such as proving the equivalency between hardware design and software model~\cite{sadeghi2021organizing}.

A recent processor fuzzer, TheHuzz~\cite{tyagi2022thehuzz}, relies on a variety of coverage metrics extracted using industrial-standard tools such as Cadence~\cite{CadenceSim} and ModelSim~\cite{modelsim}.
TheHuzz proposes an optimization strategy to increase the effectiveness of fuzzing in discovering bugs.
Specifically, TheHuzz profiles individual instructions and determines optimum instruction and mutation pairs while generating new set of inputs.
This way, TheHuzz associates individual instructions with relevant mutation strategies and aims to guide fuzzing towards buggy processor states. 
Unlike \difuz{} or \profuzz{}, TheHuzz does not propose a new coverage metric .
TheHuzz relies on several coverage metrics used in software testing (i.e., statement, branch, line, expression).
As discussed by prior works~\cite{tasiran2001functional, hur2021difuzzrtl}, these metrics are not sufficient metrics to verify a processor. 
Besides, D-flip flop (DFF) toggle coverage misses certain states as detailed by \difuz{}. Finally, it is not clear how registers that control FSM coverage are identified as the industrial-tools are not open-sourced.
Moreover, the runtime overhead of TheHuzz is higher (71\%) than \profuzz{} due to the instrumentation applied by industrial tools and profiling coverage. 
We could not quantatively compare \profuzz{} with TheHuzz as TheHuzz is not open sourced.

The common goal of the aforementioned fuzzing works is to maximize coverage of an RTL design, thereby discovering bugs across the entire RTL design.
Researchers have also proposed fuzzing frameworks for achieving alternate verification goals.
For instance, DirectFuzz~\cite{canakci2021directfuzz} adapts the notion of directed greybox fuzzing and applies it to the RTL verification. 
Contrary to the aforementioned common goal of fuzzing, the goal of DirectFuzz is to cover certain specific RTL regions with a targeted fuzzing approach.
Here, the motivation is to dedicate more fuzzing time to the RTL components that need to undergo thorough testing. 
HYPERFUZZER~\cite{muduli2020hyperfuzzing} introduces a new grammar that represents the hardware security properties. 
During fuzzing, HYPERFUZZER checks if any of the fuzzer-generated inputs violates a security property.
Defining properties requires human expertise which is error-prone.
Logic Fuzzer~\cite{kabylkas2021effective} randomizes control signals and states of a DUT without compromising the functional correctness of the DUT.
Logic Fuzzer needs to be provided with fuzzing targets (e.g., congestible points in an RTL design), and therefore requires domain expertise. 
INTROSPECTRE~\cite{ghaniyoun2021introspectre} and Osiris~\cite{osiris274675} use blackbox fuzzing approach to discover microarchitectural side channels (i.e., Meltdown~\cite{meltdown} and Spectre~\cite{spectre}) in processors.

\subsection{Differential Testing in Software Domain}
\label{sec:relatedworkdiff}

Differential testing is commonly used in the software domain to discover inconsistencies (e.g., semantic bugs, side-channels, consensus bugs) across multiple programs with similar functionalities.
One use case of differential testing in the software domain is to identify discrepancies between emulators and real hardware.
Prior works~\cite{sahin2018proteus,martignoni2009testing,paleari2009fistful} aim to eliminate the source of discrepancies in emulation environments since adversaries use discrepancies to infer the execution environment and bypass malware analysis. 
\profuzz{} differs from these works in two ways. 
First, \profuzz{}'s test input generation is coverage-guided whereas these works employ blackbox fuzzing. 
Second, these works aim to identify discrepancies that may or may not necessarily translate to actual bugs.
Besides emulators, differential testing has been used to test different types of software including Web application firewalls~\cite{argyros2016sfadiff}, SSL/TLS libraries~\cite{petsios2017nezha,brubaker2014using,chen2015guided,sivakorn2017hvlearn}, compilers~\cite{yang2011finding}, cryptocurrency protocols~\cite{yang2021finding,fu2019evmfuzzer,evmlab}, deep learning systems~\cite{pei2017deepxplore,noller2020hydiff}, Java Virtual Machines~\cite{chen2016coverage,chen2019deep}, PDF viewers~\cite{petsios2017nezha}, mobile applications~\cite{jung2008privacy}, file systems~\cite{min2015cross}, and Java programs~\cite{nilizadeh2019diffuzz,noller2020hydiff}.
\section{Discussion and Limitations} \label{sec:discussion}
\textbf{Other ISAs.} In this work, we demonstrated the capability of \profuzz{} using the RISC-V ISA. 
However, CSRs are not only specific to the RISC-V architecture and  defined as part of many other ISAs including x86.
Therefore, \profuzz{} is not limited to the RISC-V-based processors and can be used in processors based on other ISAs.

\noindent\textbf{Unintended RTL Transitions.} 
\profuzz{} uses ISA simulation as part of a feedback mechanism since it is faster and agnostic to the HDL. 
\profuzz{} does not use an input for RTL simulation if the input lacks of a unique transition in its ISA simulation trace.
One limitation of this design choice is that \profuzz{} can potentially miss certain bugs that follow the given scenario.
If a test input would result in an unintended transition in RTL simulation but the same test input does not cause any unique transition in ISA simulation, such a test input will be discarded.
Hence, the bug will not be identified.

\noindent\textbf{No RTL Coverage.} \profuzz{} does not collect feedback (i.e., CSR transitions) from the RTL design during fuzzing. 
However, we can extend its design and collect coverage from the RTL during RTL simulation to further aid fuzzing.
Note that \profuzz{} is able to discover all bugs found by prior work in its current form without using any feedback from the RTL design during fuzzing. 

\section{Conclusion}

This work presents \profuzz{}, a processor fuzzer guided by a novel CSR-transition coverage feedback obtained from ISA simulation.
\profuzz{} demonstrates that monitoring CSR transitions can effectively guide fuzzing towards buggy processor states. 
Moreover, using ISA simulation instead of RTL simulation can quickly eliminate inputs that result in the same coverage, thereby helping the fuzzer to test as many qualitatively different inputs as possible.
Our experimental results discovered eight new bugs in established, real-world, RISC-V processors, and one new bug in a reference model.

% \begin{acks}
% \end{acks}

\bibliographystyle{ACM-Reference-Format}
\bibliography{ProFuzz-CCS}

\appendix

\section{Selected and Excluded CSRs}
In this section, we provide the reasoning for the CSR selection in the \profuzz{} implementation for RISC-V ISA. 
We used the criteria mentioned in \ref{csr-selection} to select the CSRs.
In general, we selected status CSRs under first criteria (C1) and CSRs that change the configuration of the processor under the second criteria (C2). 
In Table \ref{tab:csrlist}, we show the selected CSRs along with the criteria that was used to select them.  

We also provide a list of the CSRs that are implemented in the RISC-V cores, but excluded from the selection in Table~\ref{tab:csrexcludelist}.
Exclusion of CSRs is done based on three intuitive reasons. 
First, we exclude any CSR that holds the same value throughout all tests.
For example, we maintain the same physical memory protection (PMP) configuration for all tests.
Therefore, we exclude the CSRs that configure PMP (\texttt{pmpcfg} and \texttt{pmpaddr}) because they are not expected to cause CSR transitions. 
We exclude \texttt{misa}, \texttt{mhartid}, \texttt{mtvec}, \texttt{satp} and \texttt{stvec} CSRs with the same reasoning.

Second, we exclude any CSRs that are not supported by the testing infrastructure. 
For instance, RISC-V ISA vector extension is not supported in our current instruction generator. 
Intuitively, vector extension CSRs (\texttt{vstart}, \texttt{vxsat} and \texttt{vxrm}) are not expected to cause any transitions when the vector instructions are not present. 
Hence, we exclude any CSRs from vector extension. 
Similarly, debug extension CSRs and CSRs related to handling interrupts are excluded.

Third, we exclude CSRs that does not directly represent the architectural state of the processor. 
These registers contain information to assist designers during analysis of a hardware bug rather than revealing the fundamental issue.
For example, hardware performance-monitoring counters (HPCs) provide information for hardware to assist several debugging use cases including performance bottlenecks.
Similarly, CSRs that assist in context switching and trap handling (\texttt{tval}, \texttt{scratch} and \texttt{epc} CSRs) are excluded because they similarly do not reveal the origin of a bug.
Also, note that we already monitor the trap cause CSRs (\texttt{mcause}, \texttt{scause}) and \texttt{mstatus} CSR to capture any changes in the architectural state due to exceptions and context switches.

Four, we exclude CSR that is already a subset of a CSR that we are already monitoring. For example, \texttt{sstatus} CSR is excluded because it is a subset of \textit{mstatus} CSR. 

\begin{table*}[]
\small
\centering
\caption{CSR selection for RISC-V ISA implementation of \profuzz{} along with the criteria that was used to select them. Here, C1 and C2 correspond to two criteria that we describe in Section~\ref{csr-selection}.}
\label{tab:csrlist}
\begin{tabular}{|c|l|l|c|}
\hline
\begin{tabular}[c]{@{}c@{}}\textbf{CSR}\\ \textbf{Group}\end{tabular} & \multicolumn{1}{c|}{\textbf{CSR}}          & \multicolumn{1}{c|}{\textbf{Description}}   &
\multicolumn{1}{c|}{\textbf{Criteria}}  \\ \hline \hline
                                                       & mstatus.xIE                       & Controls the global interrupt enable bit for privilege x, x = \{M, S, U\} & C2                    \\ \cline{2-4} 
                                                       & mstatus.xPIE                      & Holds the value of interrupt-enable bit active prior to the trap for privilege mode x  & C1       \\ \cline{2-4} 
                                                       & mstatus.xPP                       & Holds the previous privilege mode active prior to a trap taken to privilege mode x & C1           \\  \cline{2-4} 
                                                       & mstatus.XS                        & Contains the state of any additional user-mode extensions & C1\\
                                                       \cline{2-4}
                                                           & mstatus.FS                        & Contains the state of the floating-point unit   & C1                       \\ \cline{2-4} 
                                                       & mstatus.MPRV                      & Controls the privilege mode in which the memory operations are performed   & C2                   \\ \cline{2-4} 
                                                       & mstatus.SUM                       & Controls the permission for accessing user memory from supervisor mode & C2                        \\ \cline{2-4} 
                                                       & mstatus.MXR                       & Controls the privilege with which loads access virtual memory   & C2                              \\ \cline{2-4} 
                                                       & mstatus.TVM                       & Controls the ability to edit virtual-memory configuration from supervisor mode   & C2             \\ \cline{2-4} 
                                                       & mstatus.TW & Controls the privilege modes that wait for interrupt (WFI) is allowed to execute   & C2           \\ \cline{2-4} 
                                                       & mstatus.TSR                       & Provides the ability to trigger a trap when SRET instruction is executed in supervisor mode  & C2 \\ \cline{2-4} 
                                                       & mstatus.xXL                       & Controls the width of an integer register for privilege mode x, x = \{S, U\}   & C2               \\ \cline{2-4} 
                                                       & mstatus.SD & Indicate the combined state of mstatus.FS and mstatus.XS for context switches  & C1               \\ \cline{2-4} 
                                                       & mcause                            & Contains the trap cause when a trap is taken in to machine mode   & C1                            \\ \cline{2-4} 
                                                       & scause                            & Contains the trap cause when a trap is taken in to supervisor mode   & C1                         \\ \cline{2-4} 
                                                       & medeleg                           & Decides what type of exceptions are delegated to supervisor mode from machine mode   & C2         \\ \cline{2-4} 
                                                       & mcounteren                        & Controls the availability of the hardware performance-monitoring counters for supervisor mode & C2\\ \cline{2-4} 
\multirow{-18}{*}{Privileged}                           & scounteren                        & Controls the availability of the hardware performance-monitoring counters for user mode  & C2     \\ \hline
                                                   
                                                       & frm                               & Controls the dynamic rounding mode for floating-point operations   & C2                           \\ \cline{2-4} 
\multirow{-3}{*}{Unprivileged}                           & fflags                            & Holds the accrued exceptions from the floating-point operations  & C1                             \\ \hline
\end{tabular}
\end{table*}
\begin{table*}[]
\small
\centering
\caption{CSRs not monitored by \profuzz{} along with the reason for exclusion.}
\label{tab:csrexcludelist}
\begin{tabular}{|l|l|l|l|}
\hline
\textbf{Category}                                                                   & \textbf{CSR}           & \textbf{Description}                                                                                  & \begin{tabular}[c]{@{}c@{}}\textbf{Reason for}\\ \textbf{Exclusion}\end{tabular} \\ \hline \hline
\multirow{6}{*}{Privileged}                                                & sstatus       & Holds the supervisor mode operating status of the processor                                  & Subset of mstatus                                                     \\ \cline{2-4} 
                                                                           & misa          & Reports the CPU capabilities of a hart                                                       & \multirow{7}{2cm}{Holds a constant value during testing}                \\ \cline{2-3}
                                                                           & mhartid       & Contains the integer ID of the hardware thread running the code                              &                                                                       \\ \cline{2-3}
                                                                           & mtvec         & Contains the trap handler base address and vector configuration for machine mode             &                                                                       \\ \cline{2-3}
                                                                           & satp          & Controls supervisor-mode address translation and protection                                  &                                                                       \\ \cline{2-3}
                                                                           & stvec         & Contains the trap handler base address and vector configuration for supervisor mode          &                                                                       \\ \cline{1-3}
\multirow{2}{*}{PMP}                                                       & pmpcfg        & Contains the physical memory protection configuration                                        &                                                                       \\ \cline{2-3}
                                                                           & pmpaddr       & Contains the physical memory protection addresses                                            &                                                                       \\ \hline
\multirow{5}{*}{Interrupt}                                                 & mip           & Reports pendng interrupts in machine mode                                                    & \multirow{13}{2cm}{Not supported by the testing infrastructure}         \\ \cline{2-3}
                                                                           & mie           & Control what interrupts are enabled in machine mode                                          &                                                                       \\ \cline{2-3}
                                                                           & mideleg       & Decides what type of interrupts are delegated from machine mode to supervisor mode           &                                                                       \\ \cline{2-3}
                                                                           & sie           & Reports pendng interrupts in supervisor mode                                                 &                                                                       \\ \cline{2-3}
                                                                           & sip           & Control what interrupts are enabled in supervisor mode                                       &                                                                       \\ \cline{1-3}
\multirow{5}{1.7cm}{Debug Extension}                                           & dcsr          & Contains the configuration and status of debug extension                                     &                                                                       \\ \cline{2-3}
                                          & dpc           & Holds the program counter of the next instruction to be executed before entering debug mode  &                                                                       \\ \cline{2-3}
                                                                           & dscratch      & Optional scratch register that holds temporary values                                        &                                                                       \\ \cline{2-3}
                                                                           & tselect       & Control which trigger is accessible through the other trigger registers                      &                                                                       \\ \cline{2-3}
                                                                           & tdata1-3        & Holds trigger-specific data                                                                  &                                                                       \\ 
                                                                            \cline{1-3}
\multirow{3}{1.7cm}{Vector Extension}                                          & vstart        & Holds the index of the first element to be executed by a vector instruction                  &                                                                       \\ \cline{2-3}
                                                                           & vxsat         & Holds the saturation flag for fixed-point operations                                         &                                                                       \\ \cline{2-3}
                                                                           & vxrm          & Controls the rounding mode used in the vector extension                                      &                                                                       \\ \hline
\multirow{5}{*}{HPC}                                                       & mcountinhibit & Controls which hardware performance-monitoring counters are allowed to increment             & \multirow{11}{2cm}{Does not directly reveal the origin of a potential bug} \\ \cline{2-3}
                                                                           & cycle         & Holds the elapsed cycle count of the CPU                                                     &                                                                       \\ \cline{2-3}
                                                                           & instret       & Holds the number of retired instruction count                                                &                                                                       \\ \cline{2-3}
                                                                           & hpmevent      & Hardware performance-monitoring event selector                                               &                                                                       \\ \cline{2-3}
                                                                           & hpmcounter    & Performance-monitoring counter of the event selected by hpmevent                             &                                                                       \\ \cline{1-3}
\multirow{6}{1.7cm}{Privileged (assisting trap handling and context switches)} & mtval         & Hold the exception-specific information when a trap is taken to machine mode                 &                                                                       \\ \cline{2-3}
                                                                           & mscratch      & Holds a pointer to the machine mode context space while the hart executes in lower privilege &                                                                       \\ \cline{2-3}
                                                                           & mepc          & Contains the program counter of an instruction that caused an exception in machine mode      &                                                                       \\ \cline{2-3}
                                                                           & stval         & Hold the exception-specific information when a trap is taken to supervisor mode              &                                                                       \\ \cline{2-3}
                                                                           & sscratch      & Holds a pointer to the supervisor mode context space while the hart executes in user mode    &                                                                       \\ \cline{2-3}
                                                                           & sepc          & Contains the program counter of an instruction that caused an exception in supervisor mode   &                                                                       \\ \hline
\end{tabular}
\end{table*}

% \section{Coverage Progress for Industry-Standard Metrics}\label{appendix-covprogress}
% Here, we provide the coverage progress of \profuzz{} and \difuz{} over time for four different industry-standard coverage metrics (i.e., line, toggle, FSM, branch).

\end{document}